\documentclass[useAMS,usenatbib,usedcolumn,usegraphicx]{mn2e}
\usepackage{xspace}
\usepackage{color,epsfig,rotating,amssymb,longtable}
\usepackage{array,colortbl}
\usepackage[colorlinks=true,linkcolor=magenta,citecolor=blue,breaklinks=true]{hyperref}
\usepackage{ifpdf}

\usepackage{natbib}
\usepackage{aas_macros}

\ifpdf
\usepackage{epstopdf}%,pdfpages,pdflscape}
\else
\usepackage[dvips]{graphicx}
\usepackage{lscape}
\fi

\newcommand{\OIII}{[O{\footnotesize III}]$\lambda$5007}  
  
\newcommand{\OI}{[O{\footnotesize I}]$\lambda$6300}  
\newcommand{\SII}{[S{\footnotesize II}]$\lambda\lambda$6717,6731}  
\newcommand{\Ha}{H$\alpha$\xspace}  
\newcommand{\Hb}{H$\beta$\xspace}  
\newcommand{\Hg}{H$\gamma$} 
\newcommand{\Hd}{H$\delta$}

 \newcommand{\HII}{H\footnotesize II}
 \newcommand{\HeII}{He{\footnotesize II}$\lambda$4686} 
 \newcommand{\ojo}{\fbox{\bf !`$\odot$j$\odot$!}}  

\newcommand{\verde}[1]{{{\color{green} #1}}\xspace}

\definecolor{dgreen}{rgb}{0,.5,.1}

\newcommand{\kms}{ km\ s$^{-1}$}                            % km s-1  
\newcommand{\kmseq}{{\rm km}\:{\rm s}^{-1}}                  % km s-1  (eqn) 
\newcommand{\ergs}{erg s$^{-1}$}                            %erg s-1  
\def\ergseq{{\rm erg}\: {\rm s}^{-1}\:}                 %erg s-1  (eqn) 
\newcommand{\Msolar}{M$_{\odot}$}                           % Msolar   
\def\Msolareq{{\rm M}_{\odot}}                           % Msolar(eqn)   
\newcommand{\Lsolar}{L$_{\odot}$}                           % Lsolar   
\title[FeII absorptions]{High velocity blue-shifted FeII absorption in the dwarf star-forming galaxy PHL293B: Evidence for a wind driven supershell?}
\author[R.Terlevich et al.]{Roberto Terlevich$^{1,2}$\thanks{E-mail:
rjt@inaoep.mx}, Elena Terlevich$^{1}$, Guillermo Bosch $^{3,4}$,
 \'Angeles D\'\i az$^{5}$,   
\newauthor
Guillermo H\"agele$^{3,4}$, M\'onica Cardaci$^{3,4}$, Ver\'onica Firpo$^{3,6}$ \\
$^{1}$INAOE,  Luis Enrique Erro 1, Tonantzintla, Puebla,  C.P.~72840, M\'exico\\
$^{2}$Institute of Astronomy, Madingley Rd., Cambridge, CB3 0HA, UK\\
$^{3}$Instituto de Astrof\'isica de La Plata (CONICET-UNLP), Argentina.\\
$^{4}$Facultad de Cs.\ Astron\'omicas y Geof\'isicas, UNLP, Argentina. \\
$^{5}$Departamento de F\'\i sica Te\'orica,  Universidad Aut\'onoma de Madrid, Cantoblanco, E-28049 Madrid, Spain.\\
$^{6}$Universidad de La Serena, Chile.}
\begin{document}

\date{v 04-04 ---- Compiled   on \today\  }

\pagerange{\pageref{firstpage}--\pageref{lastpage}} \pubyear{2014}

\maketitle

\label{firstpage}

\begin{abstract}

{\it X-shooter\/} and ISIS WHT spectra of the starforming galaxy {\it PHL~293B\/} also known as {\it A2228-00\/} and {\it SDSS J223036.79-000636.9\/} are presented in this paper.

We find broad (FWHM = 1000km/s) and very broad (FWZI = 4000km/s) components in the Balmer lines, narrow absorption components in the Balmer series blueshifted by 800km/s, previously undetected FeII multiplet (42) absorptions also blueshifted by 800km/s, IR CaII triplet stellar absorptions consistent with [Fe/H] $<  -2.0 $ and no broad components or blushifted absorptions in the HeI lines. Based on historical records, we found no optical variability at the 5 $\sigma$ level of 0.02 mag between 2005 and 2013 and no optical variability at the level of 0.1mag for the past 24 years.

The lack of variability rules out transient phenomena like luminous blue variables  or SN~IIn as the origin of the blue shifted absorptions of HI and FeII.
The evidence points to either a young and dense expanding supershell  or a stationary cooling wind, in both cases driven by the young cluster wind. 

\end{abstract}

\begin{keywords}
galaxies: abundances -- galaxies: dwarfs
\end{keywords}

\section{Introduction}

Early studies of emission line galaxies have shown that objects that spectroscopically resembled HII regions both in line emission intensities and widths, constituted about 80 per cent of total samples, the rest being galaxies of Seyfert type  \citep[see e.g.][and references therein]{French80}. Some  of these objects are also characterised by their compacticity  and blue excess as shown on photographic plates and represent  the overlap between blue compact galaxies (BCG)  and  HII galaxies.  {\it PHL~293 B} is one of these objects. It was found by \citet{Haro62} in the ``Palomar-Haro-Luyten''  survey of faint blue objects.  \citet{Kinman65} obtained its optical spectrum and described it as having a  faint continuum with unresolved emission in the Balmer series and [OIII] $\lambda\lambda$5007,4959 and 4363\AA\ that makes of it, to our knowledge, one of the first detections of the [OIII]$\lambda$4363\AA\ auroral line in a star forming galaxy. We nowadays recognize this fact as the one allowing a trustworthy derivation of the gas elemental abundances in regions of star formation showing an emission line spectrum. In the case of {\it PHL~293 B}, these abundances are amongst the lowest known, less than one tenth of the solar value \citep{French80,Izotov07,Asplund09} and in the border line of what is considered to be an extremely metal deficient galaxy \citep[e.g.][]{KO00}. The galaxy was included in the sample studied by \citet{French80} who lists its absolute magnitude and size as -13.6 and 0.2 kpc respectively. {\it PHL~293B\/} also looks compact in the images obtained by \citet{Cairos01}, and \citet{Geha06} give an effective radius for this galaxy of only 0.4 kpc. Its absolute magnitude according to the Sloan Digital Sky Survey (SDSS)  is, M$_g$ = -14.77. These facts  make of it  a very low luminosity and compact HII galaxy. 

One of the characteristics of HII galaxies is their high star formation rate that takes place in a very small volume and probably in short duration episodes, thus making these galaxies easily observable. Given the large value of the equivalent width of their emission lines, it is the current burst of star formation that dominates their luminosity  at blue and visible wavelengths. The low metallicity of these objects guarantees that they are in a chemically unevolved stage probably similar to what is expected in galaxies at early cosmological times. The evolution of their massive young stars, responsible for the gas ionisation,  is conditioned by their low metallicity as is probably the case for Pop III stars. The evolution of high mass stars is short and encompasses phases with intense episodes of mass loss, therefore the presence of low intensity broad components or wings in the otherwise narrow emission lines typical of star-forming regions (SFR), are relatively common in high S/N  medium resolution spectra. Well known examples are: NGC~604, a giant HII region in the spiral galaxy M~33 \citep{Diaz87,Terlevich96}; NGC~5471, a giant HII region in the spiral galaxy M~101 \citep{Castaneda90}, NGC~2363, a giant HII region in the irregular galaxy NGC~2366 \citep{Roy92,GonzalezDelgado94} and 30~Dor in the LMC, one of the largest extragalactic HII regions in the local universe \citep{Melnick99}. This is also the case, as expected,  for HII galaxies and strong line BCG  \citep[see e.g.][]{Izotov96}, since it is the dominant star forming region that dominates their integrated spectra. 

The most likely origin of the extended line wings observed in SFR is related to the evolution of very massive stars and the feedback processes between them and the surrounding interstellar medium. Given the large number of these stars present in a relatively small volume, it may be expected that the combination of their powerful winds gave rise to complex kinematical components in both permitted and forbidden lines. Also multiple supernova events have been invoked as the possible cause \citep[see][for the case of NGC~2363]{Roy92}. However, some of these massive stars like  Luminous Blue Variables (LBV),  can be so luminous as to be capable of producing visible effects by themselves particularly in nearby resolved system and some of them may even end their lives as supernovae hence dominating for some time the galaxy luminosity output. This evolutionary path may be the one followed by the most massive stars of low metallicity for which the inefficiency of line-driven stellar winds would translate in a low mass-loss rate and failure in becoming a WR star. Some of these stars might be the progenitors of type IIn supernovae in which the narrow line spectrum arises from the interaction of the supernova blast wave with the circumstellar shell. These supernovae can be very luminous and also very long lived (Aretxaga et al. 1999; Smith et al. 2007).\nocite{Aretxaga99}

Another mechanism accounting for the presence of low intensity broad line components in HII galaxies or BCG  spectra may be mass accretion onto an intermediate-mass black hole (10$^3$ - 10$^5$ M$_{\sun}$). Up to now there are no {\em bona fide} low metallicity objects (about one tenth of solar)  harbouring massive black holes. In fact, it is intrinsically difficult to identify these objects in commonly used diagnostic diagrams \citep[e.g.][]{Stasinska06}. The overlapping of their broad component H$\alpha$ luminosities and those of supernovae and even stellar winds, makes this identification even harder. Yet, from the point of view of galactic evolution to firmly establish the existence of these objects would be of the greatest importance.
In this sense the detection of X rays emission would be a major discriminant.

In this work we analyse recently obtained moderate to high resolution spectra of {\it PHL~293B\/} which is a low luminosity, low metallicity HII galaxy that shows low intensity broad wings and blue shifted narrow absorptions in the hydrogen recombination lines, in order to try to shed some light on their origin. \S2 gives a description of the data, \S3 presents the results obtained from them and \S4 is devoted to their analysis. A discussion is presented in \S5 and the conclusions of our work are given in \S6.

\section{Description of the data}
{\it PHL~293B\/}, also known as HL~293B, Kinman's dwarf, {\it A2228-00\/} and  {\it SDSS J223036.79-000636.9\/} is a very low luminosity galaxy \citep[M$_B$ = -14.37;][]{Cairos01} at a distance of 23.1 Mpc obtained from the radial velocity \citep{Mould00} taken from NASA/IPAC Extragalactic Database (NED) corrected for Virgo Infall, Great Attractor and Shapley,  with a Hubble constant of 73 km s$^{-1}$ Mpc$^{-1}$. Its metallicity was first measured by \citet{French80} who gave a value of 12+log(O/H) = 7.78 and, more recently, by \citet{Izotov07} who derive a value of 12+log(O/H) = 7.66 from SDSS data.

\subsection{Spectroscopic data}
We have compiled spectroscopic data of {\it PHL~293B\/} for 4 different epochs spanning more than 10 years. Archival spectroscopic data have been extracted from DR7 of the Sloan Digital Sky Survey \citep{Abazajian09}, {\it SDSS\/}(J223036.79-000636.9); ESO  {\it VLT-UVES\/} [Programme ID 70.B-0717(A)]; Science Verification (SV) {\it 
VLT-X-shooter\/} [ESO program 60.A-9442(A)]. To these data we have added new observations obtained with the ISIS spectrograph attached to the 4.2m WHT at the Observatorio del Roque de los Muchachos in the island of La Palma (Spain). 

The data from SDSS were taken on the 22nd of August  2001 and  cover the spectral range from 3800 to 9200\AA\ at a resolution R = 1800-2200. The UVES data were obtained on the 18th of November 2002 in both the blue and red arms covering a total wavelength range from 3100 to 6800 \AA\ with 0.2 \AA\ spectral resolution. {\it X-shooter\/} \citep{Vernet11}  observations were performed during Science Verification on the nights of the 16th August and 28th September 2009. The data simultaneously cover the spectral range from UV($\sim 3000$ \AA ) to K$^\prime$ ($\sim$\hbox{2.5\,$\umu$m}) and were reduced using the ESO Recipe Execution Tool (EsoRex) following standard procedures with minor adaptations for these SV data. 

ISIS data were obtained on the 30th of November 2011 using an EEV12 detector attached to the blue arm of the
spectrograph. The R300B grating was used  covering the unvignetted wavelength range 4800-7200 \AA\  (centred at $\lambda_c$ = 6000 \AA ), giving a spectral dispersion of 0.86 \AA /pixel which combined with a slit width of 1 arcsec yields a resolution of about  3.4 \AA . The observations were made at paralactic angle, at an airmass of 1.3 and with a seeing of 0.5 arcsec.
Several bias and sky flat-field frames were taken at the beginning and at the end of the night. In addition, two lamp flat-fields, before and after the observation, and one calibration lamp exposure were performed. The calibration lamp used was CuNe+CuAr. The spectra were processed and analysed with IRAF routines\footnote{IRAF: the Image Reduction and Analysis Facility is distributed by  the National Optical Astronomy Observatories, which is operated by the
  Association of Universities for Research in Astronomy, Inc. (AURA) under
  cooperative agreement with the National Science Foundation (NSF).} following standard procedures that include removal of cosmic rays, bias subtraction, division by a normalized flat-field and wavelength calibration. In the last step, the spectra were corrected for atmospheric extinction and flux calibrated.  Four standard star observations were used: BD+17~4708, Wolf1346, G191B2B and Feige34 allowing a good spectrophotometric calibration with an estimated accuracy of about 5 per cent.

\subsection{Photometric data}
The photometric optical data was compiled from 3 different sources: the Catalina Sky Survey (CSS) second release \citep{Drake09} provides for {\it PHL~293B\/}, accurate  magnitudes for 80 nights from April 2005 to October 2012. Most nights have 4 different observations inside one hour total observing time. In addition, the SDSS provides photometric parameters  for the night of 22/08/2001, and \citet{Cairos01} published their photometry for {\it PHL~293B\/} obtained in October 1988 with the 3.5m telescope of the Calar Alto Observatory (Spain). \citet{Kinman65} provides the earliest brightness estimates based on the Palomar Sky Survey (PSS)  plates  and photographic images taken with the Lick  Observatory 20-inch Astrograph in 1949 and 1965 and at the prime focus of  the Lick 120-inch telescope.

{\it PHL~293B} was observed also using the ACIS-S camera on board the Chandra X-rays observatory in 2009 september 25th. Data cover the energy range 0.4 - 10\,keV and the exposure time was about 7.7 ks (ObsID 11294).

\section{Results}

\subsection{Spectroscopy}

\begin{figure}
 % \vspace*{154pt}
   \includegraphics[angle=0,width=0.5\textwidth]{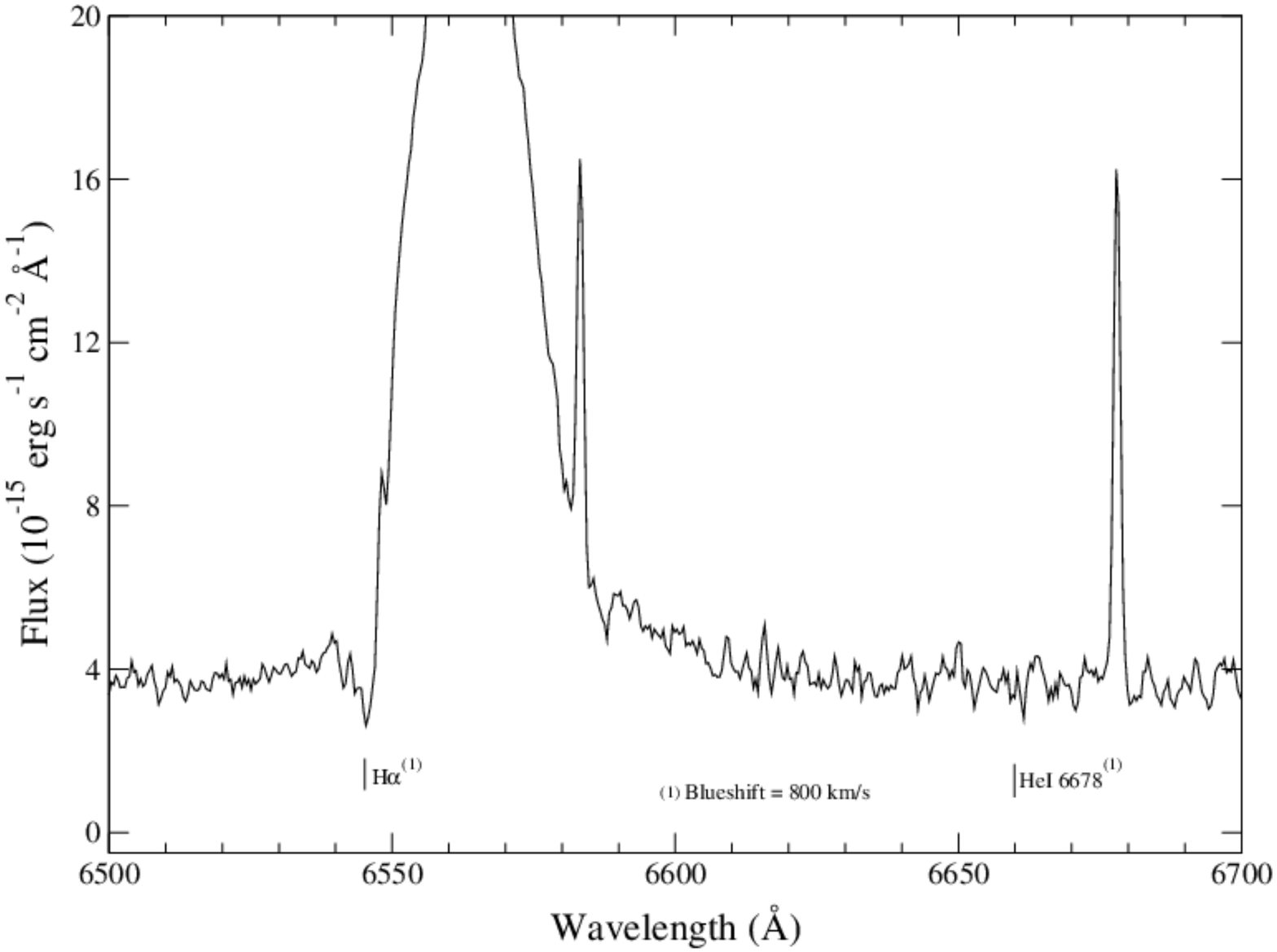}
  \caption{{\it X-shooter\/} spectrum showing the line profile around H$\alpha$. Here and in the following spectra, flux is measured in erg\,s$^{-1}$\,cm$^{-2}$.}
  \label{fig:XSH_Halpha}
\end{figure}

\begin{figure}
 % \vspace*{154pt}
   \includegraphics[angle=0,width=0.5\textwidth]{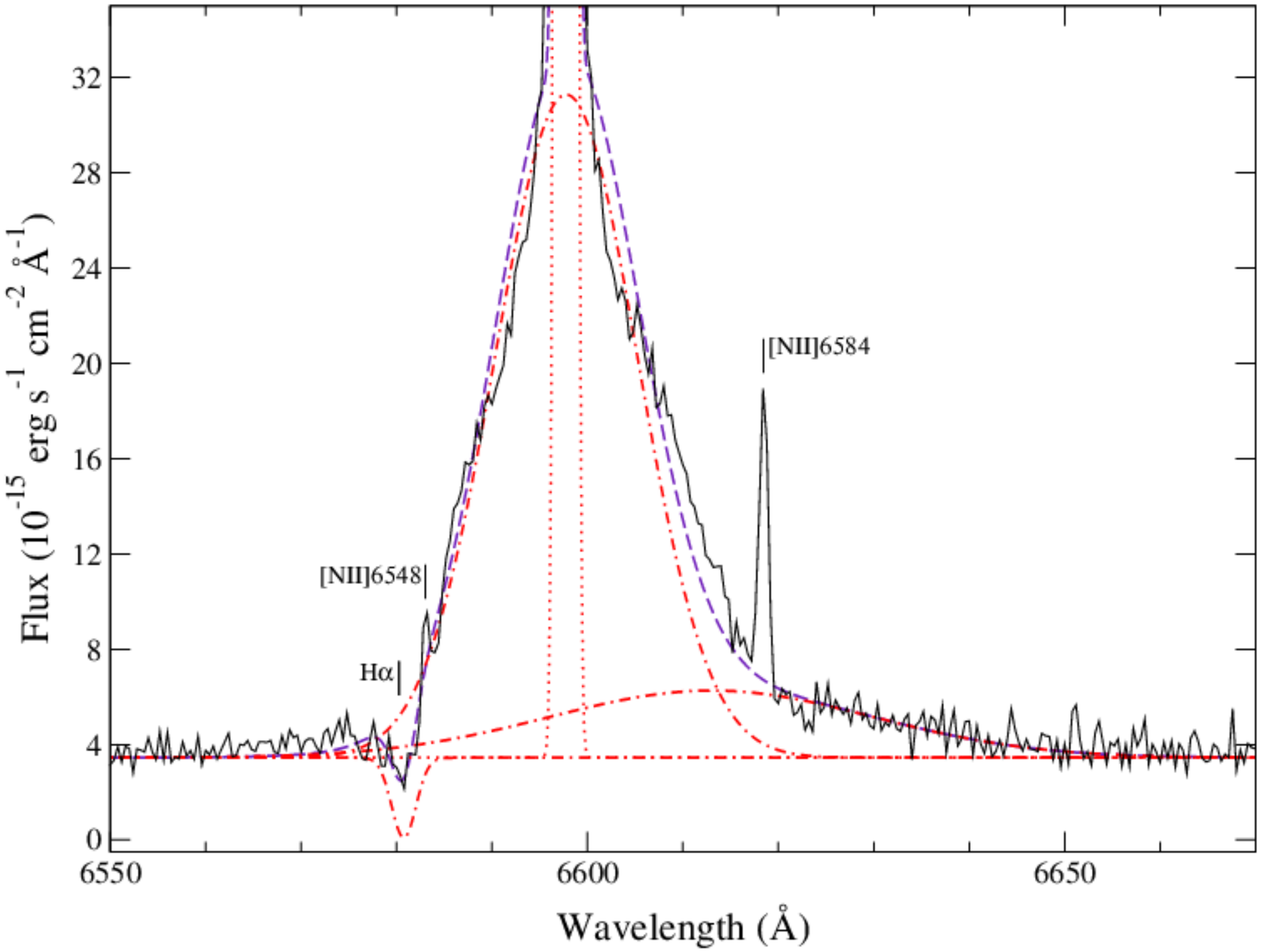}
  \caption{{\it X-shooter\/} spectrum showing the multi Gaussian fit performed to the H$\alpha$ line profile. Four components are clearly identified: a blue wing absorption blueshifted by 800 \kms, a strong narrow component, a strong broad component and a relatively fainter ultra broad component, redshifted by 900 \kms. The narrow spike at $\lambda$6620 \AA\ is an artifact from the reduction process and discarded during profile fitting.
}
  
 \label{HalphaFit}
\end{figure}

\begin{figure}
 % \vspace*{174pt}
   \includegraphics[angle=0,width=0.5\textwidth]{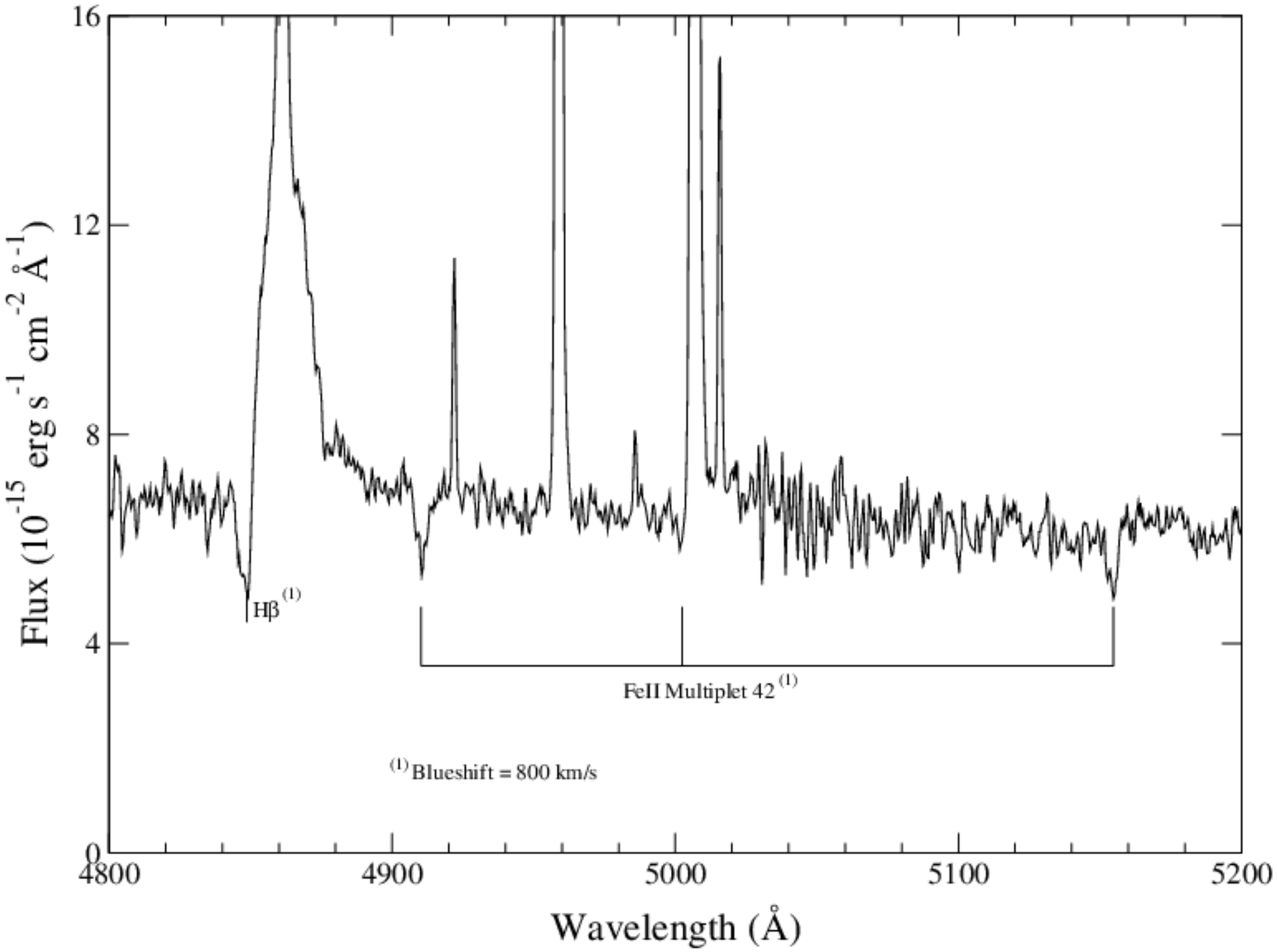}
  \caption{Same as Fig.\ 1, for the H$\beta$, [OIII] $\lambda$5007\AA\  region. Note the possibly very broad blue wing in [OIII]. The position of the FeII multiplet 42 blueshifted  by 800 \kms with respect to the galaxy rest frame is shown.} \label{Hbeta}
\end{figure}

\begin{figure}
%  \vspace*{174pt}
   \includegraphics[angle=0,width=0.5\textwidth]{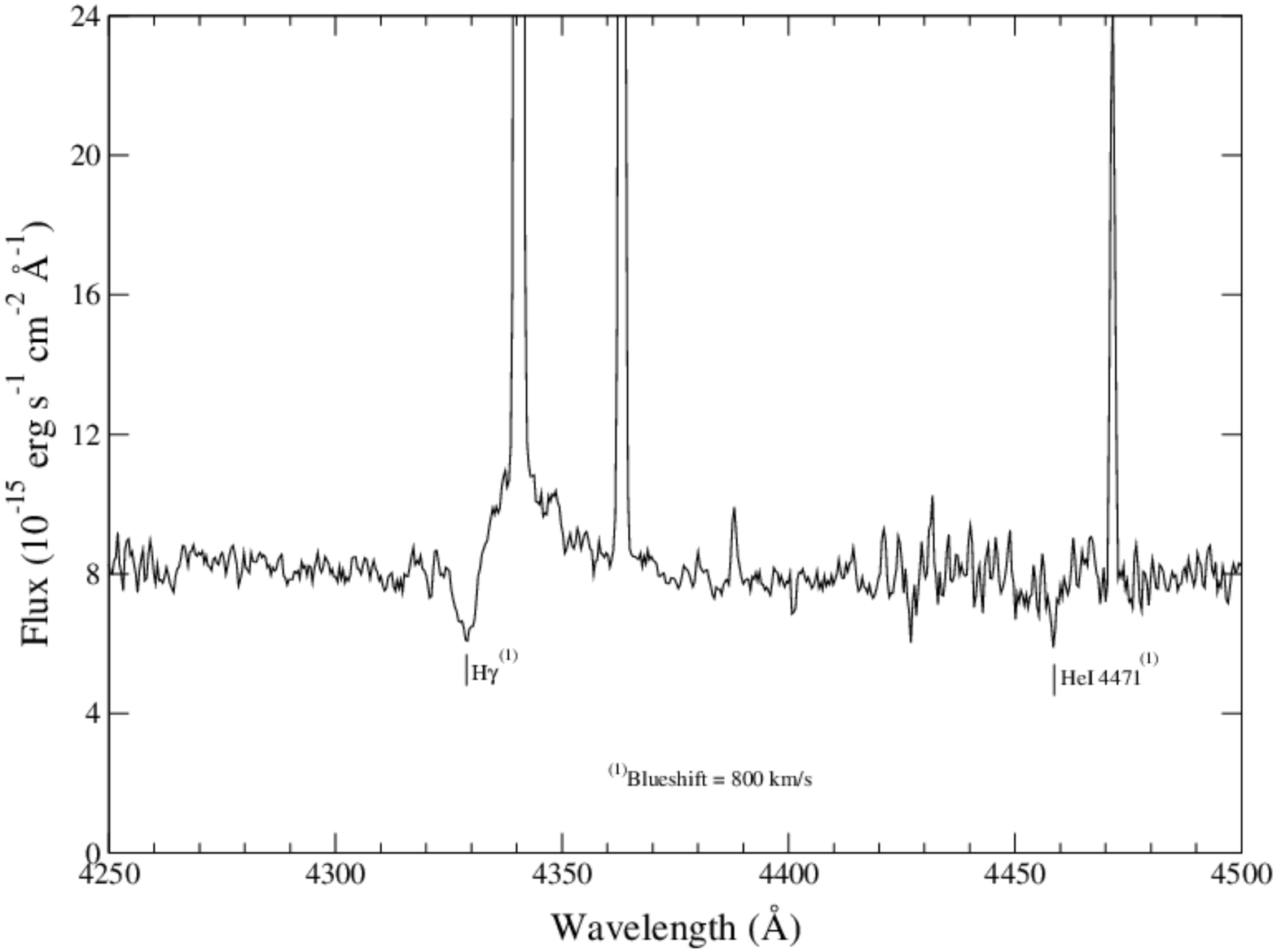}
  \caption{As Fig.\ 1, for \Hg . The positions of \Hg\/  and HeI  $\lambda$4471\AA\ blueshifted  by 800 \kms with respect to the galaxy rest frame are shown.}
  
  \label{Hgamma}
\end{figure}

\begin{figure}
 % \vspace*{174pt}
   \includegraphics[angle=0,width=0.5\textwidth]{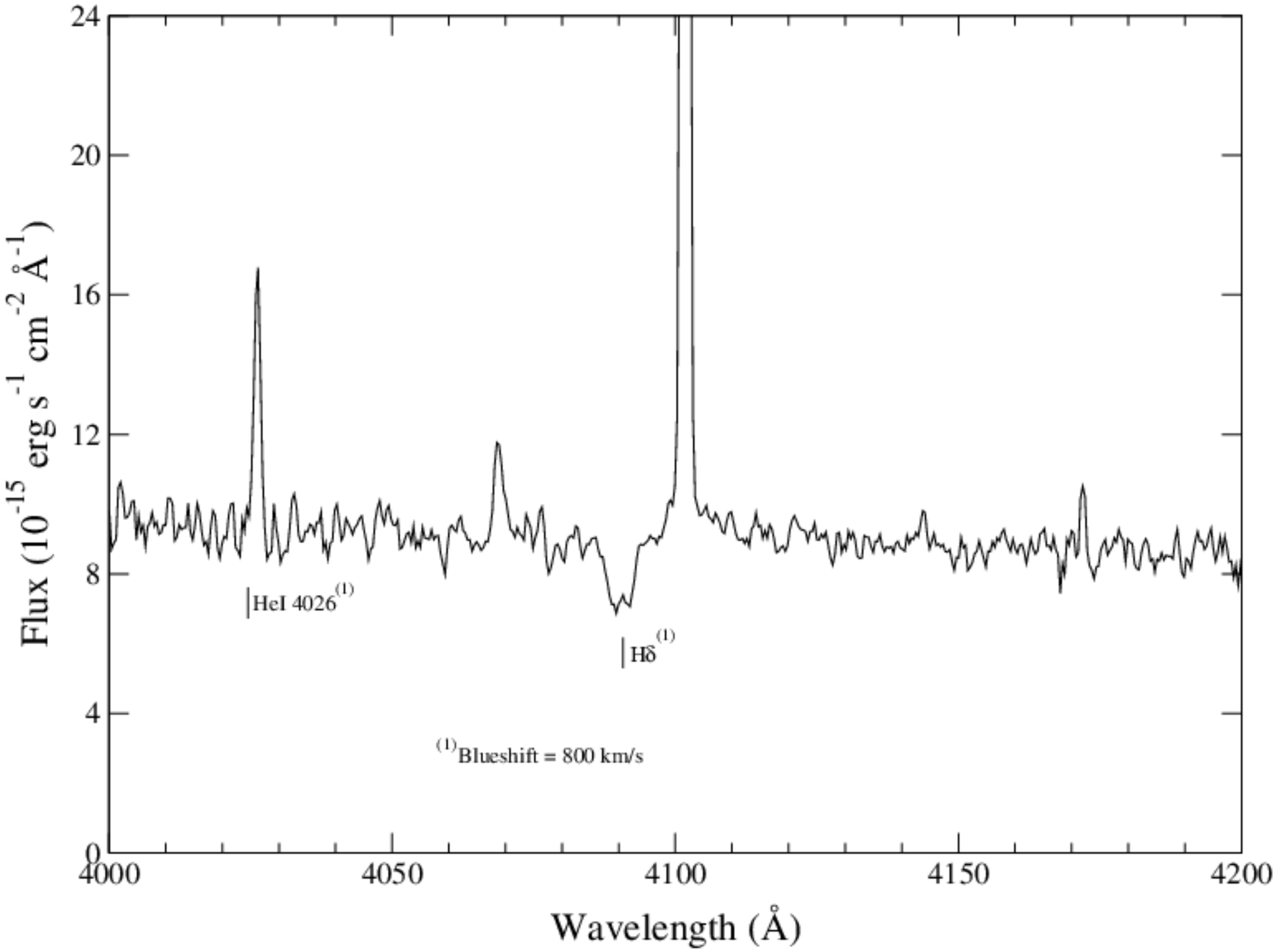}
  \caption{As Figure 1, for \Hd . The positions of \Hd\/ and HeI  $\lambda$ 4026 \AA\ blueshifted  by 800 \kms\ with respect to the galaxy rest frame are shown.} 
  
  \label{fig:XSH_Hgamma}
\end{figure}

The blue to visual spectrum of {\it PHL~293B\/} shows the usual strong emission lines found in star-forming galaxies: recombination lines from hydrogen and helium and collisionally excited lines of different ions: [OII], [OIII], [NII], [SII], [SIII], etc. The most striking feature of the spectral lines is the presence of broad components in the hydrogen recombination lines, already reported by Izotov et al. (2007), 
which are undetected in the corresponding helium lines. These broad components that can be seen in Figures \ref{fig:XSH_Halpha}  through \ref{fig:XSH_Hgamma} for the Balmer lines: \Ha , \Hb , \Hg\ and \Hd, have FWHM between 1000 and 400 km/s \citep{Izotov09,Izotov11} and  relatively low intensities. They are redshifted by about 50 km/s with respect to the narrow emission lines reference frame as determined  by the centroid of the [OIII] $\lambda$ 4959 \AA\ line. Figure \ref{HalphaFit} shows a detail of the multigaussian fit performed on the {\it X-shooter\/} spectrum 
H$\alpha$ profile.
Details of measurements performed in SDSS and {\it X-shooter\/} spectra are listed in Tables \ref{SDSSspec} and \ref{XSHspec}, using the normal notation for equivalent widths as positive for lines in absorption, negative for emission.
In what follows, we will call the reference frame defined by the [OIII] $\lambda$ 4959 \AA\ line, the galaxy reference frame. 
The fluxes in the narrow component of  H$\alpha$ and H$\beta$ (measured from the SDSS spectrum) are 4.86 and 1.54 $\times$ 10$^{-14}$ erg s$^{-1}$ cm$^{-2}$ respectively  and the ratio of broad to narrow components in both lines is about 0.25. 

 Errors in measurements depend strongly on flux uncertainties. Individual errors are determined from profile fitting routines. These errors rely on a good determination of errors present in the spectra, which is directly obtained for the SDSS spectra and had to be modeled for the {\it X-shooter\/} data. They are below 1\%\  for radial velocity and 10\%\ for dispersion determinations in strong narrow line profiles but can raise to tens of \kms for broader components. Uncertainties in flux range from 3\% in strong lines and/or components but can grow up to 20\% in weak absorption lines. Equivalent widths share a few percent to 20\% uncertainty based on relatively low counts level for the continuum flux.

From the {\it X-shooter\/} data the measured narrow line H$\alpha$ and H$\beta$ fluxes are $2$ and $ 0.66 \times 10^{-14} $ erg s$^{-1}$ cm$^{-2}$, respectively. The ratio \Ha/\Hb between the narrow line components is the same in both spectra within the observational errors. The ratio between the narrow and broad components for these lines are also the same within the observational errors. Hence the differences in the flux measurements between the SDSS and the {\it X-shooter\/} spectra can be attributed to aperture effects. Since the aperture used for the SDSS observations is larger than the one used for the {\it X-shooter\/} one, we hereafter use the SDSS measurements for the line fluxes and EWs.
Additional features of the Balmer lines are an ultra-broad component with FWZI $\sim$ 4000Km/s redshifted by about 500 km/s  and a P~Cygni like absorption blueshifted by 800 km/s with respect to the reference frame (see table \ref{XSHspec}).

\label{Emission line profiles}
To summarise,  the complex \Ha\ line profile shows a narrow component at the galaxy reference frame,
a broad component FWHM $\sim$ 1000 km/s redshifted by about 50 km/s with respect to the galaxy reference frame,
an ultra-broad component with FWZI $\sim$ 4000 km/s redshifted by about 500 km/s and
a P~Cygni like narrow absorption blueshifted by 800 km/s.

These features are also seen in the other Balmer lines, except for the ultra-broad component which is seen only as a red wing. A weak red wing is also visible in [OIII]$\lambda$5007\AA\ corresponding possibly to the ultra-broad component. All these features are visible both in the {\it X-shooter\/} data analysed here and in the SDSS spectrum. The broad component represents about 20\% of the total H$\alpha$ emision flux. The blue shifted absorption  is weaker in 
H$\alpha$ than in the other Balmer lines, although the continuum is difficult to fit in the blue wing of H$\alpha$. 

The HeI lines show only narrow components.

%%%%%%%%%%%%%%%%% FIGURE 1  %%%%%%%%%%%%

%%%%%%%%%%%%%%%%% FIGURE 2  %%%%%%%%%%%%

%%%%%%%%%%%%%%%%% FIGURE 3  %%%%%%%%%%%%

%%%%%%%%%%%%%%%%% FIGURE 4 %%%%%%%%%%%%

\begin{figure*}
 \includegraphics[angle=0,width=1.00 \textwidth]{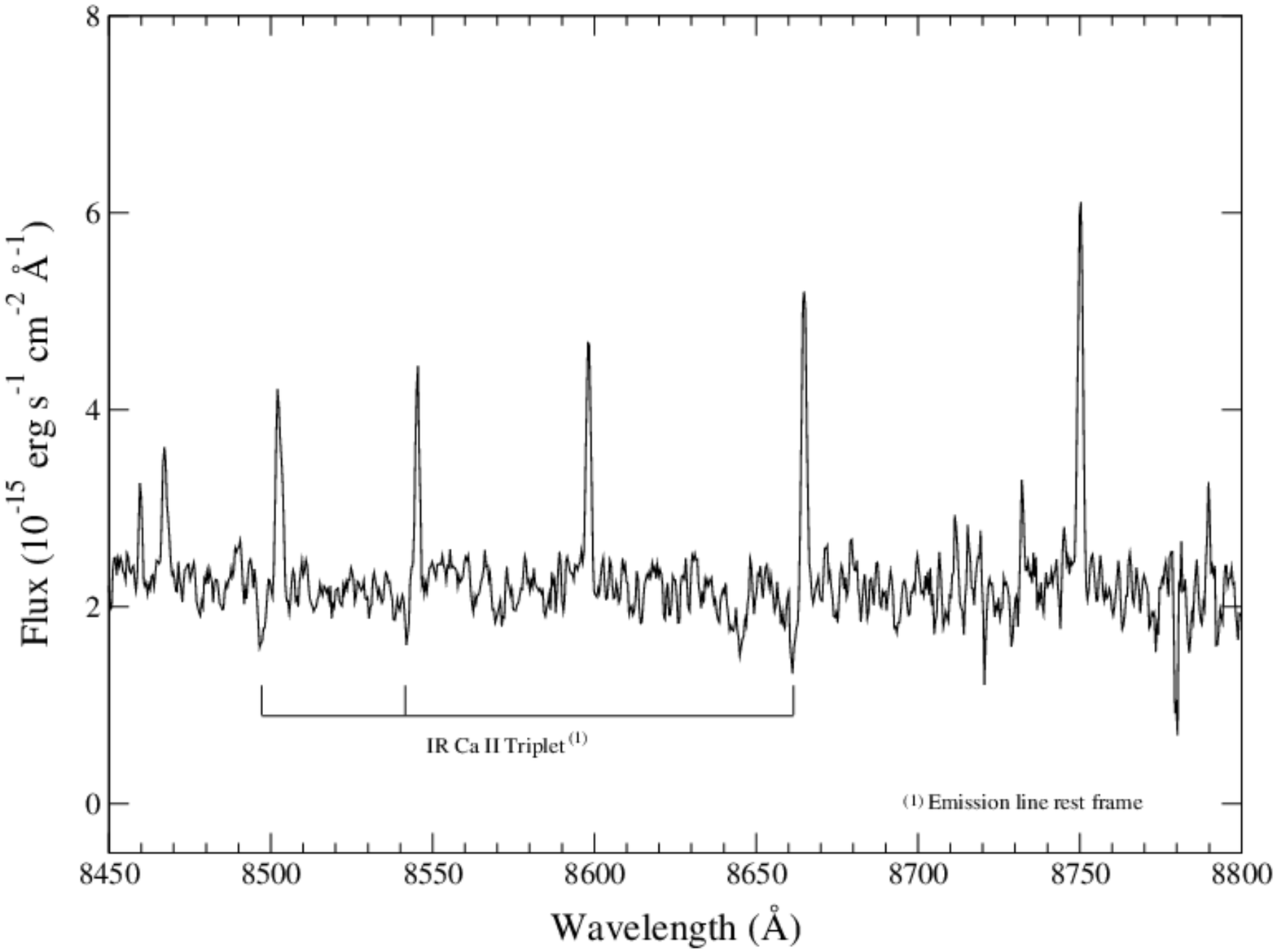}
\caption{Visible-NIR range of the  {\it X-shooter\/} spectrum  showing the emission lines from the Paschen series and the IR CaII triplet stellar absorptions. } \label{CaTspec}
\end{figure*}

We have identified a conspicuous absorption feature at $\lambda$ 5183 \AA . This absorption line has no emission counterpart and coincides with the wavelength of the FeII multiplet 42 (laboratory wavelength  5169.03 \AA ) line blueshifted by the same 800 km/s displayed by the Balmer P~Cygni like components. Companion lines of the same multiplet blueshifted by the same velocity have also been found at  $\lambda\lambda$ 5031, 4939 \AA\ and they are indicated in Figure \ref{Hbeta}. 
Further absorption lines are observed in the far red spectral region, corresponding to the CaII triplet (CaT) lines
at $\lambda\lambda$ 8498, 8542, 8662 \AA\ in the galaxy rest frame (see Figure \ref{CaTspec}). Although these lines, being close to the left of the adjacent Paschen lines at $\lambda\lambda$ 8505, 8548, 8667 \AA ,   can mimic P~Cygni profiles, the CaII lines are clearly separated from the Paschen ones. Furthermore, there is no hint of a P~Cygni like component in the $\lambda$ 8601 \AA\ Paschen line.

Although the {\it X-shooter\/} infrared spectrum is not a very good one, we can identify hydrogen recombination lines of the Paschen and Brackett series, HeI lines and a hint of the H$_2$ line at 2.122 $\mu$m rest frame. There is no evidence for broad components or  P~Cygni like profiles in any of these lines. 

\medskip

\begin{figure*}
 % \vspace*{174pt}
   \includegraphics[angle=0,width=1.0 \textwidth]{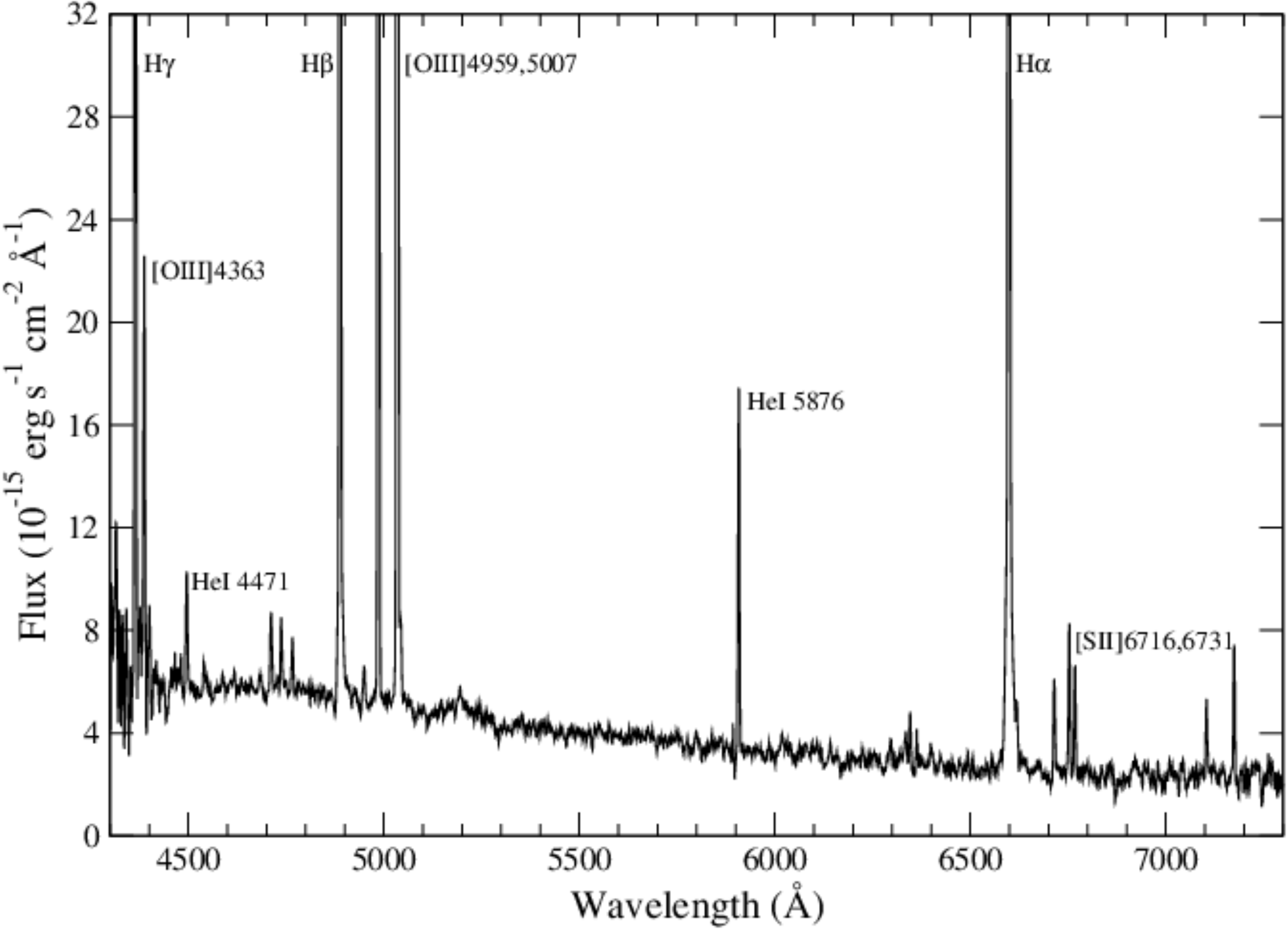}
  \caption{WHT-ISIS spectrum box smoothed by 5 pixels.}\label{ISISv}
\end{figure*}

The WHT-ISIS spectrum covering from $\lambda$4300 \AA\ to $\lambda$7500 \AA\ (observations set-up described in \S2.1) is shown in Figure \ref{ISISv}.
The strongest emission lines are labelled.

%%%%%%%%%%%%%%%%% FIGURE 5 %%%%%%%%%%%%
%%%%%%%%%%%%%%%%%%%%%%%%%%%%%%%%%%%%%%

%%%%%%%%% Figure 6 Far red  spectrum %%%%%%%%%%%%%%%

%%%%%%%%%%%%%%%%%%%%%%%%%%%%%%%%%%%%%%%%%%%%

%%%%%%%%%%%%%%%% Table SDSS %%%%%%%%%%%%%%%
\begin{table*}
\centering
\caption{SDSS spectroscopy: Column 1 lists the line identification and its measured wavelength is included in Column 2. Derived radial velocities are shown in Column 3 and observed velocity dispersion, corrected by instrumental broadening, are listed in Column 4.  Errors in Columns 2 to 4 are shown in parentheses, in \kms . Also, fluxes and equivalent widths  (positive in absorption, negative in emission) of line profiles are shown in Columns 5 and 6 respectively. Relative errors, in percentage, are shown in parentheses. Note that no errors are shown for the FeII(42)5018 line because the close proximity of the strong He\,I line prevented the iterative fitting procedure.}

\label{SDSSspec}
\begin{tabular}{cccccc}
\hline
Id.\       &  $\lambda_{\mathrm{obs}}$ & Radial Vel.\ & $\sigma_{\mathrm{obs}}$ & Flux                                                & EW \\
            &  (\AA  )                                  & \kms          & \kms                                & 10$^{14}$ erg s$^{-1}$ cm$^{-2}$ & (\AA  ) \\
\hline
FeII(42) 1 $\lambda$4924     & 4939.6 (0.5)  & 955 (31)  & 133 (33)   &        & 1.0 (23\%)  \\
FeII(42) 2 $\lambda$5018     & 5030.3          & 710:         &  56:           &       & 0.5:          \\
FeII(42) 3 $\lambda$5169     & 5183.0 (0.4)  & 810 (20)  &   33 (20)    &       & 0.67 (23\%)    \\
\Hg                                        & 4364.6 (0.02) & 1665 (1) & 59 (2)     & 0.7 (1.5\%) & -36 (1.5\%)   \\
\Hb (narrow)                           & 4888.4 (0.01) & 1672 (1) & 25 (1)    & 1.5 (1.2\%) &  -82 (1.2 \%) \\
\Hb (broad)                            & 4890.4 (0.5)  &  1793 (30) & 321 (34)  & 0.3 (8\%) & -21 (8\%) \\
\Hb (absorption)                    & 4874.6 (0.3)   & 818 (17)  & 302 (19)   &                 & 1.1 (16\%) \\
\Ha  (narrow)                          & 6599.5 (0.01) & 1678 (1) & 36  (1)   &   4.7 (0.5\%) & -489 (0.5\%) \\
\Ha  (broad)                            & 6601.2 (0.7)  & 1756 (30) & 681 (33)   &   1.3  (5\%) & -138 (5\%)  \\
\OIII                                       & 5034.8 (0.01) & 1675 (0.2) & 18.3 (0.1)  &  8.6  (1\%) & -594 (1\%) \\
\end{tabular}
\end{table*}

%%%%%%%%%%%%%%%% Table X-Shooter %%%%%%%%%%%%%%%
\begin{table*}
\centering
\caption{{\it X-shooter\/} spectroscopy: Column descriptions are similar to those in Table \ref{SDSSspec}. Column 5 lists individual component fluxes relative to the total flux measured for the overall emission profile.  The red wing component in the \Ha profile is fitted with a fix width, so the error determinations of all its components are not reliable, and therefore not included.} 
\label{XSHspec}
\begin{tabular}{cccccc}
\hline
Id.\ & $\lambda_{\mathrm{obs}}$ & Radial Vel.\ & $\sigma_{\mathrm{obs}}$  & Flux                                & EW \\
      &  (\AA  )                               & \kms               & \kms                              &  Relative to total line flux & (\AA  ) \\
\hline
\Hb (absorption)               & 4874.1 (0.1) & 789 (6)  & 130 (11)   & -0.01 & 2 (10\%)  \\
\Hb (narrow)                     & 4887.0 (0.01) & 1582 (1) & 26 (01)  & 0.84 & -96 (3\%)  \\
\Hb (broad)                       & 4887.8 (0.2) & 1629 (12) & 466 (13)  &  0.19 & -21 (2\%)  \\
\hline
FeII(42) 1 $\lambda$4924 & 4936.7 (0.2) &  777 (12) & 115 (15)   &          &  1 (12\%) \\
FeII(42) 3 $\lambda$5169 & 5181.6 (0.2) &  733 (12) & 99 (15)     &           & 1 (10\%) \\
\hline
\OIII                                  & 5033.3 (0.002) & 1588 (0.1) & 22 (0.1)  &        & -584 (1\%) \\
\hline                           
\Ha (absorption)               & 6580.8 (0.16)  & 823 (7)   & 91 (08)  & -0.01 & 6 (10\%) \\
\Ha (narrow)                     & 6597.8 (0.02) & 1599 (1) & 17 (01)  & 0.77   &  -530 (3\%) \\
%\Ha (narrow 2)                  & 6597.6 & 1590 & 32  & 0.13 &  -95   \\
\Ha (broad)                       &  6598.2 (0.2) & 1620 (9) & 394 (10) & 0.20  & -136 (3\%) \\
\Ha (red wing)                   &  6617.7 & 2510 & 172 & 0.04 & -29  \\
\hline
Ca II $\lambda$8498        & 8542.6 (0.2) & 1575 (5) & 30 (6) &          & 0.7 (17\%)  \\
Ca II $\lambda$8542        & 8587.7 (0.3) & 1604 (10) & 26 (10) &          &  0.4 (12\%) \\
Ca II $\lambda$8662        & 8707.7 (0.1) & 1577 (3) & 24 (3) &          &  0.7 (13\%) \\
\end{tabular}
\end{table*}

%%%%%%%%%%%%%%%%%%%%% Table FeII %%%%%%%%%%%%%%%
\begin{table*}
 \centering
 \begin{minipage}{160mm}
  \caption{{\it X-shooter\/}, {\it UVES\/},  {\it SDSS\/} and {\it ISIS\/} equivalent width of absorption lines of \Hb\/ , components 1 and 3 of the FeII(42) multiplet and [OIII] $\lambda$4959 \AA . Measurements  are in the galaxy rest frame.  Flux in units of $10^{-17}$ \ergs .}
  \label{FeII}
  \begin{tabular}{@{}|l|ccc|cccccc|ccc|@{}}
\multicolumn{13}{l} {}\\

  \hline
  
Spectra    &   & H$\beta$                & &&&        FeII(42)           &  &  &&       &  [OIII]4959  &\\

  &&&	&		 			     	   &     1        &&         & 3      &&  && \\

		&	Wav.&	EqW & Vr&      Wav. &  EqW &  Vr  &    Wav. &  EqW   &  Vr   &  Wav. & Flux  &   EqW \\
		&	(\AA) &	(\AA) & \kms &      (\AA). &  (\AA) &  \kms  &    (\AA). &  (\AA)   &  \kms   &  (\AA). &   &   (\AA) \\

  \hline
    \hline
    
SDSS   	&	4847.3 &1.1	& -866. & 4912.2 &1.0 &  -715.  & 5154.5& 0.67&  -843. &  4959.1& 164 & -171. \\

UVES        &4848.1 & 2.0 & -816. &----&----& -----&                     5154.7 &0.88 & -832.  & 4959.0 &192 & -208.  \\

XShooter    & 4848.0 &  1.1 & -822. & 4910.8 & 0.91&  -800. &  5154.5 & 0.68 & -843. &  4959.0 &186 & -197.  \\

ISIS      &  4847.9 & 1.0 &-829. &     4912.0      &0.71&-727&           5154.7 &0.57 &-832.  & 4959.1& 220&  -211.\\
\hline

\end{tabular}
\end{minipage}
\end{table*}

\subsection{Photometry}\label{photometry}

\begin{figure*}
\centering
\includegraphics[angle=0,height=10cm,width=16cm]{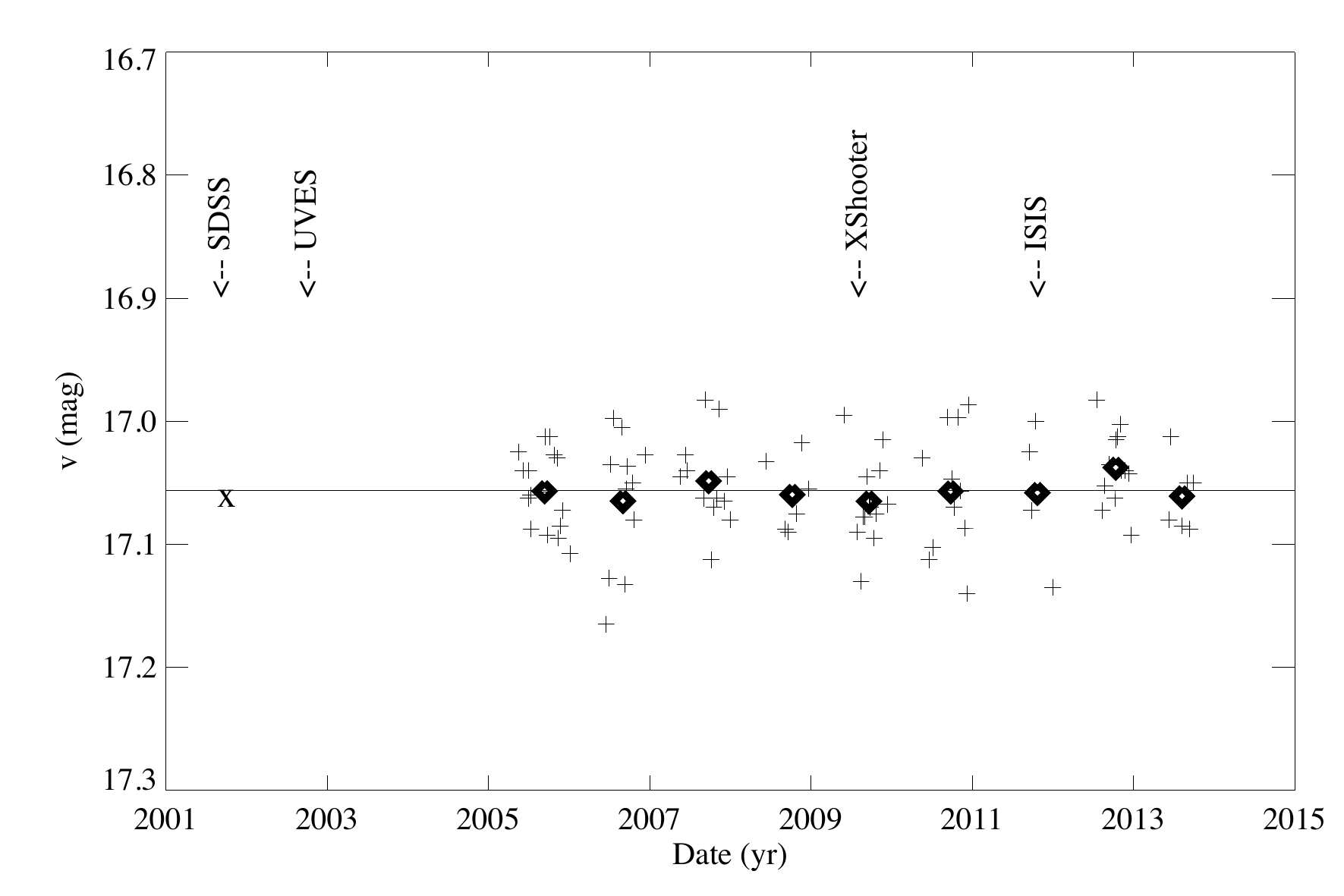}
\caption{{\it PHL~293B\/} CSS photometry from  April 2005 to October 2013. There are 353 observations in 91 observing nights. Most night results are made out of four independent observations. The average night magnitude is plotted with a plus sign. The 9 filled rhomboids are the yearly averages. The horizontal line represents the average magnitude 17.056 $\pm$ 0.004. The x indicates the SDSS Petrosian g magnitude converted to v magnitudes.  For older photometry see text. The labelled arrows show the dates corresponding to the spectroscopy.}
\label{catalina}
\end{figure*}

We have extracted from the Catalina Sky  Surveys  (CSS)  data release 2 database  \citep{Drake09} 353 photometric points for {\it PHL~293B\/} corresponding to 91 observing nights over the 8.4 years from 16/5/2005 to 27/9/2013. Most nights have 4 independent observations inside a 60 minutes total span. 
Nightly averages are shown (plus symbols) in Figure \ref{catalina} and have a typical r.m.s. scatter of 0.040 magnitudes. The yearly averages are shown with filled rhomboids. The average brightness for the whole dataset, indicated with a thin line, is
$<V> =$17.056 $\pm$ 0.004.

The SDSS  published Petrosian r magnitude  is P$_r= $17.04 $\pm $ 0.01 which combined with the colour g-r=0.03 and the SDSS colour transformations yield a Petrosian v magnitude P$_v=$17.07  $\pm $ 0.01 observed on the 22/08/2001. The SDSS value is indicated with an x in Figure \ref{catalina}.

\citet{Cairos01} provided an  earlier photometric point for the object. Their October 1988 observation with the 3.5m telescope of the Calar Alto Observatory gives V=17.02$\pm $ 0.01. 

\citet{Kinman65} published the earliest magnitude estimates  for {\it PHL~293B\/}: m$_{pg}=$16.7.  We quote from his paper: ``a very rough estimate of the B magnitude is 17.7 on the Sky Survey plates and about  a half magnitude fainter on the 120-inch plates". Kinman attributes this difference to the scale difference between the PSS and the Lick 120-inch plates given that there is no  variability between 1949 and 1965  in plates taken with the Lick 20-inch Astrograph. 
Bearing in mind that they are based on eye estimates  of the  PSS plates from 1949  and photographic images taken at the prime focus of  the Lick Observatory 120-inch telescope in 1965, errors of about $\pm $0.5 magnitudes are expected.

There is no detection of X-rays emission in the ACIS-Chandra image. The point source sensitivity of this camera in the 0.4 to 6 keV energy range is $4 \times 10^{-15}$ erg\,cm$^{-2}$\,s$^{-1}$ for an exposure time of 10 ks \citep[Table 6.1 of][]{ChandraPOG}. It gives for the {\it PHL~293B\/} Chandra observation of 7.7 ks a point source sensitivity of about $3.5 \times 10^{-15}$ erg\,cm$^{-2}$\,s$^{-1}$.

%\subsection{ HST images} We have extracted 5 images from the HST database. They correspond to WFC3 + UVIS with filters F336W,F438W, F606W and F814W. A faint star in the field was used to verify the PSF of the images. In all of them PHL was marginally resolved. After correcting for the PSF the effective radius we obtain is 2.5 parsecs.

%%%%%%%%% Figura 7 ISIS spectrum %%%%%%%%%%%%%%%%%%

%%%%%%%%%%%%%% Figura 8 VARIABILITY DATA %%%%%%%%%%%%%

%%%%%%%%%%%%%%%%%%%%%%%%%%%%%%%%%%%%%%%%%%%%%%%%

\section{Analysis}
\subsection{Ionising cluster age, mass and abundance}

{\it PHL~293B\/} is a low luminosity  \citep[M$_B$ = -14.37;][]{Cairos01}, low metallicity [12+log(O/H) = 7.61; Izotov et   al.~2007] BCG. The comparison of our measured  narrow \Ha\ to \Hb\ flux ratio with the case B recombination theoretical value yields a logarithmic extinction at \Hb\, of c(H$\beta$) = 0.12. For our SDSS flux measurements and the assumed distance of 23.1 Mpc (NED) this gives an extinction corrected H$\beta$ luminosity of 1.3 $\times$ 10$^{39}$ erg s$^{-1}$. The total Lyman continuum photon rate for a given region can be derived from the \Ha\ luminosity \citep[e.g.][]{LeithererHeckman95}:

\[ N(Lyc) = 7.35 \times 10^{11} L(H\alpha)  = 2.71 \times 10^{51} s^{-1}\]

Taking into account that the number of Lyman continuum photons per unit mass of the ionising population decreases with age, and using the equivalent width of H$\beta$ (EW(\Hb )) as an age parameter \citep{Dottori81}, it is possible to calculate the mass of the ionising star cluster under certain assumptions about the initial mass function (IMF). For a Salpeter  IMF  with  lower and upper mass limits of 0.8 and 120 M$_{\sun}$ respectively, we have \citep{Diaz00}: 

\[ log[N(Lyc)/M_{ion}]=44.48 +0.86 ~log[|EW(H\beta)|] \]
where the vertical bars indicates modulus. 

For an EW(H$\beta$) = -100 \AA , that represents the total EW of H$\beta$ (see Table \ref{SDSSspec}),  the resulting ionising star cluster mass is M$_{ion}$= 1.66 $\times$10$^5$ M$_{\sun}$.  

According to the models of \citet{Molla09},  an  EW(\Hb )~=~-100~\AA\ may be produced by a cluster of 3.43$\times$10$^5$ M$_{\sun}$ with an age of about 7$\times$10$^6$ yr, for a metallicity of Z=0.0004 or a cluster of 2.06$\times$10$^5$ M$_{\sun}$ with an age of about 5$\times$10$^6$ yr, for a metallicity of Z=0.004. The ionised gas metallicity of {\it PHL~293B\/} is about Z=0.002, i.e. in between the values  of the two models quoted, hence intermediate values for the mass of the ionising cluster and the age can be expected. The observed equivalent width may be affected by the continuum of an underlying older stellar population or a bright stellar component like for example a SN or an LBV; therefore  it has to be considered as a lower limit for the equivalent width of the ionising cluster, and our mass and age estimates are in consequence upper limits.

\subsection{CaT lines}
The  IR CaII triplet at nominal  $\lambda\lambda$ 8498, 8542, 8662 \AA\ is clearly seen in absorption in the spectrum of {\it PHL~293B\/}  (see Figure \ref{CaTspec}) at the same redshift shown by the narrow emission lines. The CaT index definition \citep{DTT,TDT}  that uses the two strongest lines of the triplet, measures approximately 2 \AA , which, if taken literally, suggests [Fe/H]$<-2.0$ for giant or supergiant stars \citep[see e.g.][]{DTT2}. 
This index has to be corrected for the contribution of the nebular continuum. For the model of log(age)=6.86 and Z=0.0004 chosen above, this contribution amounts to 0.55 \AA\ per line and hence the corrected CaT index would be 3.1 \AA , consistent with the values calculated by \citet{GarciaVargas98} from single stellar population synthesis models for the lowest abundance they consider, Z=0.004. 

For the assumed mass, age and metallicity of the ionising cluster the models by \citet{Molla092} predict no WR stars and only one or two RSG stars.

\subsection{Ionised gas mass}
{\it PHL~293B\/} is unusual as an HII galaxy due to the presence of broad permitted and narrow forbidden lines in its optical spectrum.
The absence  of  a strong broad forbidden \OIII\ component indicates that the density of the gas producing the broad component is high enough to collisionally de-excite \OIII\  (a condition for which the electron density has to be N$_e \geq 10^8 \rm{cm}^{-3}$). An upper limit to the mass of the ionised gas in the broad line region can  be estimated using the observed L(\Ha) \citep{Macchetto90} and the electron density lower limit, 

\[ M(\HII) = 3.32 \times 10^{-33} L(H\alpha)/N_e \]
therefore, 
\[M(\HII )_{\rmn broad\, comp} \lesssim 3 \times 10^{-2} \Msolareq \]

The ionized gas mass of the narrow line region is, assuming an electron density of 100 cm$^{-3}$ (derived from the [SII] doublet ratio in our ISIS data),

\[M(\HII )_{\rmn narrow\, comp} \sim 1.22 \times 10^5 \Msolareq \]
similar  to the ionising star cluster mass determined in \S4.1.

\subsection{Photometric Variability (or lack of it)} \label{photvar}

From the nightly averaged data (pluses) in Figure \ref{catalina} no variability is detected.   Dividing the data  into 9 groups of  one year duration each (filled diamonds in the same figure) reduces the r.m.s. scatter to 0.009 per group as expected for a non variable sample and again no systematic trend in the luminosity is apparent; furthermore
a cursory analysis indicates that there is no periodic variation.

Based on the CSS data we can infer that at  3 $\sigma$ level any long term variability, i.e. over a few years,  cannot be larger than 0.02 magnitudes and that  any medium term variability, i.e. over a few months, should be  less than 0.04 magnitudes.
Further still, including the data from \citet{Cairos01}, \citet{Kinman65} and SDSS we can safely conclude that there is no variability at the level of few tenths of magnitude over a period of 25 years.

\section{Discussion}

\subsection{P~Cygni profiles}

To generate relatively narrow P~Cygni like absorption profiles such as the ones observed in {\it PHL~293B\/}, the absorbing material must cover a substantial fraction of the continuum source, in this case a young stellar cluster 
 several parsecs in size. These absorptions may be produced in an expanding dense supershell created by the interaction of  combined stellar winds with the  circumcluster medium. An expanding supershell has been postulated to explain the blueshifted metal absorption detected in the UV spectrum of several nearby and high redshift HII galaxies 
\citep[]{GTT99, Mas_Hesse03}.

Alternatively the observed blueshifted narrow absorptions could be associated with a single star  in a very luminous transient phase like an LBV or a SN~IIn.

\subsection{P~Cygni profiles and the HeI lines}

The possibility that the P~Cygni like profiles in {\it PHL~293B\/} are originated in the powerful wind of an LBV has been extensively discussed by \citet{Izotov11}. These authors suggested that the absorptions seen at $\lambda$ 4939 \AA\ and  $\lambda$5031\AA\  are in fact the blue shifted absorptions associated with the HeI lines at rest wavelengths $\lambda$4921 \AA\ and $\lambda$5016 \AA\ in accordance with the P~Cygni profile of the Balmer lines. 

\begin{figure*}
\centering
\includegraphics[angle=0,width=16cm]{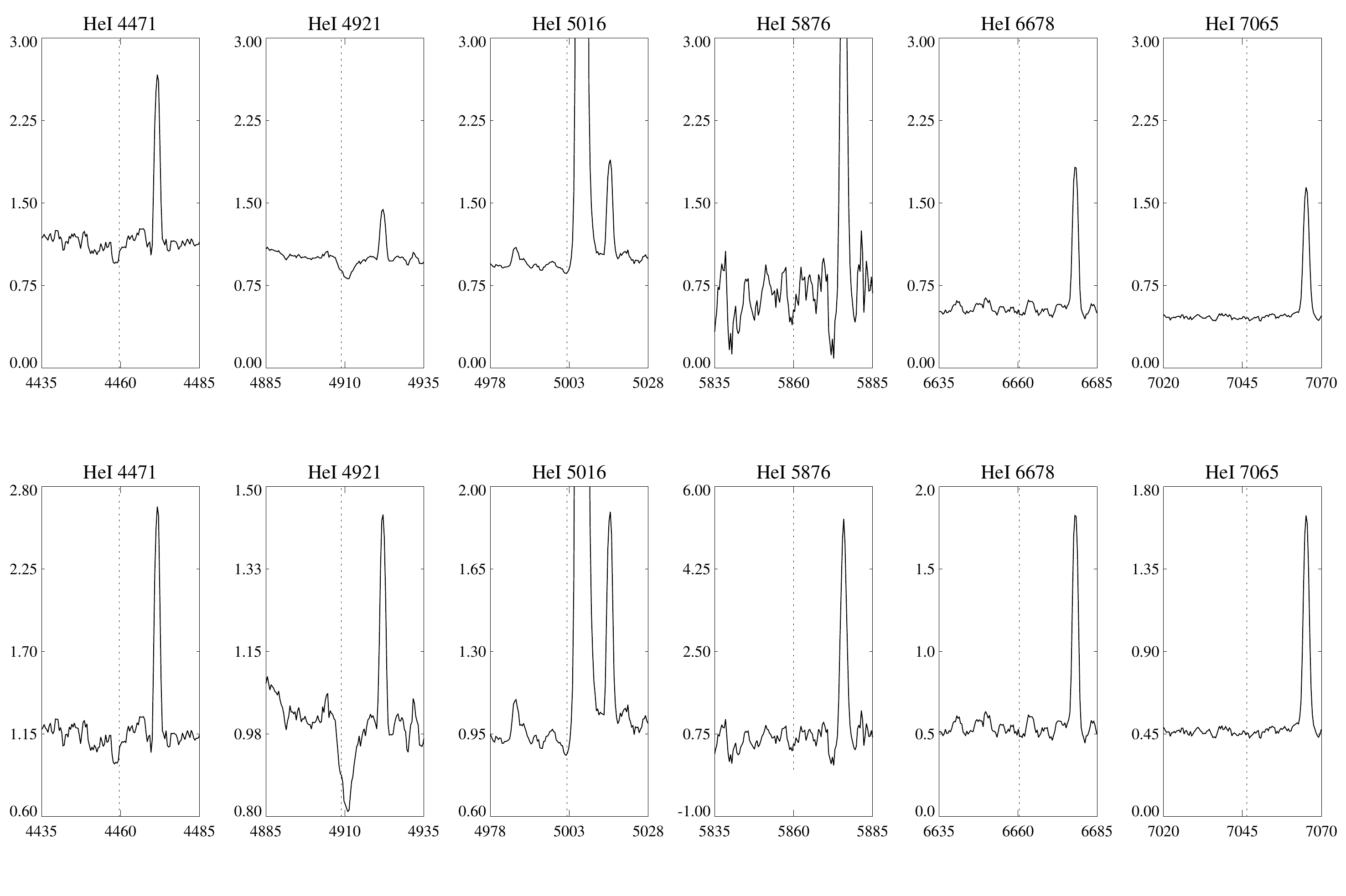}
\caption{Spectral regions around the HeI optical lines from the {\it X-shooter\/} data. The dotted line corresponds to a blueshift of 800Km/s. Top row spectra are plotted with the same scale in flux to show the relative importance of the emissions. Bottom row shows the spectra normalised to the same amplitude for the emission line and the continuum to illustrate the differences in the associated absorptions. See text regarding the strong absorption to the blue of HeI$\lambda$ 4921 \AA, shifted with respect to the 800 \kms\ line.}
\label{multiple230}
\end{figure*}

Figure \ref{multiple230} shows the spectral regions around the HeI optical lines from the {\it X-shooter\/} data. The dotted line corresponds to a blueshift of 800 \kms\ consistent with what is observed in the Balmer lines. The top row displays all the HeI  lines with the same flux scale in order to show the relative importance of the emissions. The bottom row shows the spectra normalised to the same amplitude for the emission line and the continuum to illustrate the differences in strength of the associated absorptions. 

It is clear that while the HeI $\lambda$ 4921 \AA\ emission is the weakest, the absorption to the blue is the strongest and furthermore it is redshifted with respect to the 800 \kms\ line.  
\citet{Izotov11} attributed this difference in velocity between Balmer and HeI lines P~Cygni like  profiles to the different formation depth of the core of each line in an accelerated wind. 
However there are several difficulties with this suggestion. Firstly the broad absorption to the blue of HeI $\lambda$4921 \AA\ has an equivalent width similar to that of the P~Cygni like absorption of $H\beta$, probably too large to be associated with such a weak HeI line. Secondly, it is not clear why much stronger HeI lines like $\lambda\lambda$5876, 6678 and 7065 \AA\  do not show a blueshifted  absorption as can be seen in figure \ref{multiple230}, although HeI 5876\AA\ seems to have, at the noise level, an associated  weak absorption.
 
 %%%%%%%%%%%%%%%  FIGURE 9 Regions around HeI lines  %%%%%%%%%%%%

%%%%%%%%%%%%%%%%%%%%%%%%%%%%%%%%%%%%%%%%%%%%%%%%%%%%%

\subsection{The absorption lines of FeII multiplet 42 }

\begin{figure*}
\includegraphics[angle=0,height=18cm,width=16cm]{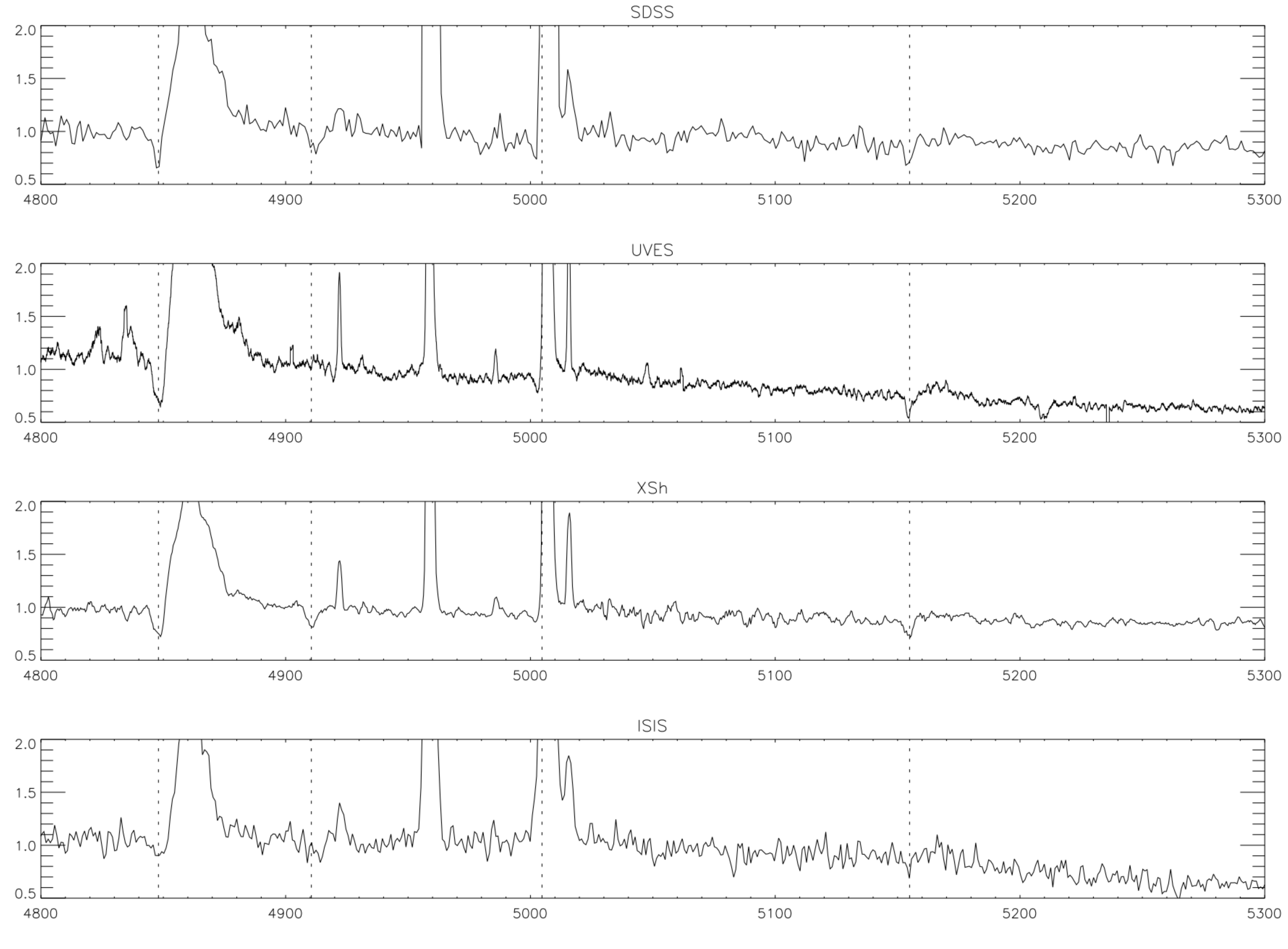}
\caption{The four panels show the SDSS, VLT-UVES, VLT-{\it X-shooter\/} and WHT-ISIS data in the range 4800-5300 \AA. 
The position of the H$\beta$ and the FeII(42) multiplet  absorptions all blue-shifted by 800  km/s,  are indicated with dotted lines. }
\label{multiple2}
\end{figure*}

An important clue for an alternative interpretation of the absorptions seen at $\lambda$ 4939 \AA\ and  $\lambda$5031 \AA\  is given by the discovery of a  previously unidentified absorption at $\lambda \sim$5183 \AA\ (Fig.\ \ref{Hbeta}) corresponding to  $\lambda$5155 \AA\  in the rest frame of the forbidden lines and  $\lambda$5169 \AA\  in the rest frame of the blue shifted Balmer absorptions.
This absorption is real and is present in the spectra from SDSS, VLT-UVES, VLT-{\it X-shooter\/} and WHT-ISIS that were taken in four different epochs (see Fig.\,\ref{multiple2}). Although the  $\lambda \sim$5183 \AA\ absorption could be attributed to Mg2 $\lambda$5172  \AA\ present in late star spectra, this possibility  can be readily dismissed given that the observed wavelength would imply a shift of 1050 \kms with respect to the galaxy rest frame and definitely inconsistent with the presence of the CaII triplet lines at the galaxy velocity rest frame.

A more plausible interpretation is that we are in fact detecting the FeII multiplet (42)  $\lambda\lambda$ 4923.93,5018.44,5169.03\AA\ blueshifted by the same amount as the Balmer absorption features  i.e.~$\sim$800 km/s to $\lambda\lambda$ 4911,5005,5155 \AA. These absorptions have already been detected in a variety of transient luminous objects like the LBV candidate NGC~2366V1 or SN~IIn like 1995G, 1999el or 1999eb among others \citep[see Figure 2 of][]{DiCarlo02}.

LBVs represent  a short ($\sim$ 10$^4$ yr)  phase in the evolution of massive stars during which the star undergoes luminous outbursts associated with large mass loss events. During these short phases lasting few years the LBV can reach, in extreme cases, 10$^6$ \Lsolar\ or Mv $\sim$ -10  and luminosity variations of $\sim$ 1 mag. The luminosity changes are in general correlated with spectral changes. The typical absolute magnitude  of an LBV  in quiescent state is  Mv $\sim$ -6. In general information about LBVs is very sparse given that only 35  candidates have been  identified in our galaxy \citep{Clark05}. Only two of these, $\eta $ Car and P~Cygni, were observed during an outburst. Extragalactic examples are known due to either photometric or specific spectral variability. The mechanism responsible for the outbursts  remains unidentified.

 It is important to mention that the wind terminal velocities quoted by \citet[][700-850 \kms]{Izotov11} are significantly larger than the usual range for LBVs (100-250 \kms \ up to  500 \kms \ for $\eta$~Car) \citep{Leitherer94}. This discrepancy is attributed by Izotov et al.~ to  {\it PHL~293B\/}  very low metallicity. 
 
There are two striking examples of LBVs in low metallicity dwarf galaxies, one in NGC~2366 and the other one in IC1613.

NGC~2366 is a giant HII region (similar to 30~Dor in the LMC) located in the metal poor \citep[12+log(O/H)=7.9, i.e.\ O/H $\sim\ $ 0.16 solar,][]{GonzalezDelgado94} dwarf galaxy NGC~2363, a member of the M81 group.
The variable star NGC~2366-V1 showed a big eruption similar to an LBV with an amplitude of 3.5 mag in 4 years
reaching a maximum of about Mv=-10.2 \citep{Drissen97}. After the outburst it started a decline of about 0.25 magnitudes in 6 years accompanied by $\sim $0.2 mag dips lasting about a year each. Regarding its spectral evolution,  the Fe II lines (4923, 5018, and 5171) present between 1997 and 2000 disappeared later while the HeI transitted from absorption to emission and the P~Cygni feature in H$\alpha$ disappeared \citep{Petit06}.
These authors reported an average wind terminal velocity of $\sim$ 300 \kms\ for NGC~2366-V1.

IC 1613 is a dwarf irregular galaxy in the Local Group with a metallicity of 
log(O/H)$+$12 = 7.80$\pm\ $0.10 
as determined from its B-supergiants \citep{Bresolin07}. Nebular abundance estimates vary between 7.60 and 7.90 \citep{Lee03}, similar to the metallicity values shown by NGC\,2363 and {\it PHL~293B\/}.

The peculiar variable star V39 \citep{Sandage71} was studied in detail by \citet{Herrero10}. Its spectrum shows strong Balmer and FeII P~Cygni profiles combined with weak HeI emission similar to the early stages of the evolution of the NGC\,2366-V1 LBV star. The equivalent width of the \Ha\ P~Cygni profile  was 
$\sim $45  \AA\ in emission and  $\sim$2.3 \AA\  in absorption with a blue shift or wind terminal velocity  of  
$\sim $400 km/s.
Its absolute magnitude reached Mv = -8.1.

The FeII (42) absorptions in these two transient objects show equivalent widths of about 2 \AA\ while the observed value in the spectrum of {\it PHL~293B\/} is just under 1\AA . The contribution of the nebular continuum at the FeII wavelengths is 25 per cent (see above), which increases this value to 1.3 \AA . This implies that under this hypothesis, about 30 percent  of the observed continuum should be due to the transient object itself. 
But, the absolute magnitude of {\it PHL~293B\/}  is Mv = -14.5 while a very luminous LBV during outburst would reach Mv $\sim\ $ -10, thus contributing  less than 2 percent of the total luminosity i.e.~more than one order of magnitude smaller than what is needed to explain the observed strength of the FeII absorptions (see Table \ref{FeII}).

It seems important to notice that these two LBVs, both located in low metallicity environments, show wind terminal velocities that are typical of LBV (100-250 \kms)  while $\eta$~Car, belonging to a region that is not metal poor
 \citep[12 + log O/H=8.36 $\pm$0.03; ][]{Pilyugin03}, has a larger wind terminal velocity (up to  500 \kms) .

Wind velocities between 100 km/s  and 1000 km/s are reported for transient events or ``supernova impostors" that could be related to powerful eruptions of LBVs \citep{Smith11}, and 
in a very interesting recent work \citet{Koss14}  report an unusual variable source in the nearby dwarf galaxy MRK 177 (UGC~239) that they suggest can be explained as an LBV eruption followed by a  SN~IIn like event in 2001. 

\citet{Naze12} discussed the X-rays properties of the Galactic LBVs. They found that their X-ray luminosities in the 0.5-8 keV range are between $\sim8\;\times\;10^{29}$ and $\sim4\;\times\;10^{34}$ erg\,s$^{-1}$. The estimated minimum luminosity for a point source to be detected in the ACIS-Chandra image of  {\it PHL~293B\/} is about $2.2 \times 10^{38}$ erg\,s$^{-1}$ (see section \S\ref{photometry}) about three orders of magnitud higher than the largest reported values for Galactic objects.

%%%%%%%%%%%%%%%  FIGURE 10 Comparison of FeII absorption from different data sets  %%%%%%%%%%%%

%%%%%%%%%%%%%%%%%%%%%%%%%%%%%%%%%%%%%%%%%%%%%%%%%%%%%%%%%%%%%%%%%%%%%%%%%%%

\subsection{Lack of optical variability}

One fact supporting the hypothesis of the presence of a transient object would be the detection of long term variability in the flux from {\it PHL~293B\/}. However, as stated in \S\ref{photvar}, we can conclude that there is no variability at the level of a few hundredths of magnitude over 25 years. This implies that any long term variable component has to be fainter than V $= $21.5. For a distance modulus  (m-M)$ = $31.8  this upper limit means that any long term variable component has to be fainter than M$_V= $-10.5. However, this does not exclude the possibility that we are observing such a transient object during a non variable phase. The caveat  in that case is 
that, if the variable component is  that faint in relation with the ionising cluster, it will be difficult to explain the observed strength of the Balmer and FeII absorptions.

\subsection{An old SN type IIn?}

An alternative scenario for the broad lines and pseudo P~Cygni profiles in {\it PHL~293B\/} is that they are originated in an evolved SN type IIn. 
 It is usually assumed that SN~IIn explode inside the circumstellar material previously ejected in the form of a slow dense wind by the red supergiant SN  progenitor. What is observed is the result of  the interaction between the ejecta and this slowly expanding  dense circumstellar medium (CSM) which transforms the mechanical energy of the ejecta into radiation. These events are known as SN~IIn or compact supernova remnants 
 \citep[see e.g.~][]{Chugai91,Terlevich92,Turatto03}.  In this interaction, radiative cooling can become catastrophic and these remnants rapidly radiate most of the mechanical energy in a relatively short time scale and so they are highly luminous. The energy output is mostly in the extreme UV and X-ray regions of the spectrum. The optical continuum is a mixture of young stars and the emission from the SN~IIn with fast moving high density gas producing the broad line emission.

The spectra of some  SN~IIn associated with  HII regions like 1995G, 1999el or 1999eb among others, present broad Balmer components with relatively narrow P~Cygni like absorptions plus narrowish FeII absorption with multiplet (42) being the strongest  \citep[see figure 2 of][]{DiCarlo02}. 
 A significant constraint on the SN~IIn/compact remnant scenario is the long term lack of variability in  {\it PHL~293B\/}. 
 The secular variability of the broad lines in this scenario comes from the evolution of the interaction of the ejecta and the circumstellar medium, i.e. the evolution of the compact remnant. The evolution of the blue light-curve can be parametrized as  $L_{B} \propto t^{-11/7}$  \citep{Aretxaga94,Aretxaga97} with its  peak luminosity  being a function of the circumstellar density (${\rmn L_{B}}^{\rmn peak} \propto n^{3/4}$). These are analytical approximations and their predicted values should be taken as such.
  
 We note that the observed absolute magnitude of  {\it PHL~293B\/} ($\sim -14.8$) is  fainter than the absolute magnitude of a single compact remnant near peak ($\sim -15.5$ for $n \sim  10^{4}$ cm$^{-3}$).
 This, together with the upper limit for the rate of change for the last $\sim\ $ 30 years, is only consistent with the very late stages of the evolution of a compact remnant, i.e.~when its light curve flattens. 
 Using the $t^{-11/7}$ time dependence of the luminosity as an indicator, we find that a compact remnant of age $\sim 150$ years evolving in a medium with density $n \sim 10^{4}$ cm$^{-3}$ would have shown a decline of about 1.5 magnitudes since maximum and  of about 0.1 magnitudes in the last 10 years.

 We can estimate its line luminosities  using the models of \citet{Terlevich92}. 
 The observed luminosity of the broad component of \Ha\ is:
 
 \[
 {\rmn L}_{H\alpha} \sim  10^{39}\ergseq
 \]

 From  the analytical solutions of \citet{Terlevich92} and tables 1.1 and 1.2 of \citet{Terlevich94} it is possible to estimate that  a compact remnant evolving in a medium with $n \sim 10^{4}$ cm$^{-3}$ would have a shock velocity of 800 km/s at an age defined by $t_{sg} $ of about 112 yr. 
Predicted values for the shock parameters and H$\alpha$ corresponding to such age  are:

  \[
{\rmn  T}_{Sh} \sim 1 \times 10^{7} K
 \]
  \[
{\rmn  L}_{Sh} \sim 1 \times 10^{41}\ergseq
 \]
 \[
 {\rmn L}_{H\alpha} \sim 3 \times 10^{39}\ergseq
 \]
 \[
{\rmn  R}_{Sh}  \sim 0.3 ~pc 
 \]
 \[
{\rmn  n}_{Sh} \sim 4 \times 10^7 cm^{-3}
 \]
 \noindent
 and the FWZI of the broad component is about twice the shock velocity or 1600 \kms . Given its temperature, the remnant is expected to  emit in X rays up to about few keV with a luminosity just below  $\sim 10^{41} $\ergs . These values are too large compared with the observed upper limit of $\sim 2.2 \times 10^{38} $\ergs .
 Furthermore, the expected drop in luminosity  7.5 years later i.e. at an age of 119.5 yr is  10 \%  or about 0.1 magnitudes and the drop in luminosity in 20 years is about 0.2 magnitudes, thus we can safely conclude that the observed upper limit on the variability of  {\it PHL~293B\/} makes it very difficult for the observed absorptions and broad emission to be originated in a SN type IIn. 
 
In summary, we have evidence that an old compact remnant could explain the observed broad lines and the dense shell responsible for the blue-shifted absorptions  in {\it PHL 293B\/}. But, while the observed line properties are in rough agreement with a compact remnant slightly older than 100 years, this age is much smaller than the age derived from the absolute blue magnitude limit imposed by the lack of variability and furthermore the expected variability  in H$\alpha$ at  100 years is also too large compared with the limits deduced from the data.

 \subsection{Spectral variability (or lack of it)  }

To see if there is some information in the data regarding the spatial distribution of the  absorbers, we compared in Table \ref{FeII} the EW measurements of the hydrogen absorptions that are supposed to cover most of the source of the continuum, with the ones of the FeII lines taking as reference the EW of \OIII\  that is expected to extend beyond most of the ionising cluster continuum but to be still inside the host galaxy continuum.

The EW of  the \Hb\ and FeII absorptions are, within  errors, basically the same in all four sets of data.

To check for aperture effects we measured the EW of \OIII\ and the results are displayed in Table \ref{FeII}.
As expected there is a slight increase in the EW measured with the smaller {\it X-shooter\/} aperture (1 arcsec) compared with the 3 arcsec circular aperture of the SDSS spectrum (although determination of faint continuum contribution can also account for such variations among spectra).

We have also checked for possible variability in the broad emission components. The  broad and narrow components of H$\alpha$ are listed in Tables  \ref{SDSSspec} and \ref{XSHspec}. Their ratio of the broad to total  H$\alpha$ flux is, for both SDSS and X-Shooter observations, the same within errors, about 0.21 showing no indication of  variability  as already discussed.

\subsection{The expanding supershell and stationary cooling wind scenarios.}

The lack of variability imposes  a strong constraint for scenarios involving discrete transient sources like LBV or SN~IIn.
At the same time the lack of variability suggests that the origin of the blue shifted absorptions could be associated with a global process involving a large part of the ionising cluster. Furthermore we have to address the fact that what we are witnessing in the FeII lines and perhaps also in the Balmer series  in  {\it PHL 293B\/} are  blue shifted absorptions without the associated narrow emissions typical of classical P~Cygni profiles.
This lack of associated emissions indicates that the material responsible for the detected absorptions, while covering a large part of the continuum source, is not  an extended circum-cluster medium (CCM). If it was it would have produced the so far undetected associated emission lines.

Luminous regions of star formation in the nearby universe  like Orion or NGC~3606 in our galaxy, 30-Doradus in the LMC or NGC~604 in M33, to name just a few of the nearest ones, all show gaseous filaments surrounding the ionising cluster. This fact and the observed lack of variability over many years suggest a scenario where the blueshifted absorptions found in  {\it PHL 293B\/} are formed in an expanding supershell generated by the cluster wind interacting with a moderately dense gas closely surrounding  the star-forming region.
The cluster wind drives a shock into the CCM  and a recombined dense shell is formed in the post-shock region. This dense shell expanding at 800 km/s produces the observed blue shifted absorptions in H and Fe following catastrophic radiative cooling in a scenario similar to the case of the compact supernova remnant described above.  The main difference being that instead of an instantaneous input of energy and mass the wind scenario involves  a constant rate. As the cluster wind shocks the surrounding ISM it will generate, given adequate conditions, a post shock high density supershell  that will be ionised by the cluster UV output modified by the radiative shock emission. 

As already mentioned, a  supershell  scenario has been postulated to explain the diversity of Ly$\alpha$ profiles observed in starbursts  \citep[see][and references therein]{GTT99,Mas_Hesse03} where also blueshifted UV ISM metal absorption lines are observed.
In this scenario the Ly$\alpha$ emission and the blueshifted metal absorption lines are formed in an expanding supershell  generated by the interaction of the combined stellar winds and supernova ejecta from the young starburst, with an extended gaseous halo. 

Among the many intrinsic parameters of a star-forming region that can affect the
properties of the observed emission line profiles, velocity, density and ionisation distributions of the gas along the line of sight
are by far the dominant ones. 
The expulsion of dust and gas from young clusters due to the action of stellar winds and  supernovae has been discussed by many authors
\citep[see for example][]{Tutukov78,Goodwin06,Bastian06} as well as the resultant cluster ``infant mortality" effect \citep[see for example][]{Grosbol13}.

\label{Can we estimate the supershell  column densities, its mass or Fe abundance?}

The strength of a weak absorption line depends on its oscillator strength $f$, its wavelength $\lambda$ and column density N. If the element is mostly in one ion state and there is no line saturation the relation between the  column density N and the EW$_{\lambda}$ is given by \citet{Morton91}:

\[ log (N)  =  log (EW_{\lambda}/\lambda) -  log (\lambda f )   + 20.053  \]

From the {\it X-shooter\/} spectra of  {\it PHL 293B\/} we have measured EW = 0.91 \AA\  and  0.68 \AA\  for the  FeII lines 4923\AA\  and 5169\AA\ respectively. The corresponding oscillator strengths are 0.0104 and 0.0226 \citep{Kramida12,Giridhar95}.
With these values for the oscillator strength we can compute the ${Fe^+}$ column density,  

\[N_{Fe^+} \sim 4.1 \times 10^{14} cm^{-2}\] and   \[N_{Fe^+} \sim 1.4 \times 10^{14} cm^{-2}\]
for the  FeII lines 4923\AA\  and 5169\AA\ respectively. 
The ${Fe^+}$ column density values are in good agreement with each other within observational errors. 

If the supershell  is formed from shocked CCM, it must have abundances similar to those observed in the ionised gas. If Fe/O is solar, this implies for the supershell   Fe/H $ \sim 3 \times 10^{-6}$ and therefore the hydrogen  column density associated with ${Fe^+}$ is,

\[N_{H^{+}} \sim 10^{20} cm^{-2}\] or \[N_{H^+} \sim 2 \times 10^{20} cm^{-2}\] if Fe/O is half solar. Given that we do not know the hydrogen ionization fraction, these values should be taken as an estimate of the total, i.e. neutral plus ionized, hydrogen column density.

Following the same procedure we can estimate from the Balmer absorptions the column density of neutral hydrogen. For a measured EW = 6 \AA\  and  2 \AA\  for \Ha\ and \Hb\  the column densities estimates 
of neutral hydrogen are respectively:

\[N_{H} \sim 3 \times 10^{13} cm^{-2}\] and  \[N_{H} \sim 8 \times 10^{13} cm^{-2}\]

The comparison of the column density of ``total'' and neutral hydrogen suggests a high degree of ionisation in the gas that is producing the blueshifted absorptions.

An interesting variant of this scenario is provided by the Super Star Cluster (SSC) cooling wind model of \citet{Silich04}, \citet{GTT07} and collaborators. 
This group has shown that in the case of a very massive and extremely compact young cluster its wind may radiatively cool in a stationary condition close to the outer radius of the cluster. This cool ejecta should be the responsible agent for the broad emission and blue shifted absorptions observed \citep[see figure 4a of ][]{GTT07}). We expect to see soon detailed model calculations for the supershell  and stationary cooling wind scenarios and the comparison of the theoretical predictions with the observed parameters of  {\it PHL~293B\/}.

If the FeII absorptions are formed in a stationary cooling wind the metal abundances will be those of the combined ejecta of stellar winds and SNe. In this scenario Fe/H will be much higher than that of the ISM in {\it PHL 293B\/} and consequently the estimated $N_{H^+}$ column density will be proportionally lower than those estimated above. Assuming solar abundances for the cluster wind, the estimated ionized hydrogen  column density will be: 

\[N_{H^+} \sim 10^{19} cm^{-2}\]

A  question that is raised in these relatively long lived cluster wind scenarios is:  why is it that among several thousand HII galaxies known, only {\it PHL~293B\/} is known to have simultanously narrow blue shifted Balmer and FeII absorptions? or equivalently, why is this type of event so rare?

The answer may be  related to the fact that we are witnessing  an event that produces weak  narrow absorptions that are detectable only with high dispersion and high S/N spectra.

\section{CONCLUSIONS}

We have analysed spectra of the low metallicity starforming galaxy {\it PHL 293B\/} 
corresponding to four epochs obtained in four different combinations of telescopes and spectrographs, the SDSS, VLT-UVES, VLT-{\it X-shooter\/} and WHT-ISIS. 

We find moderate narrow absorption components in the Balmer series blueshifted by 800km/s.
We detected also the IR CaII triplet lines at the galaxy velocity rest frame, i.e. the rest frame defined by the ionised gas narrow emission lines.
We also find narrow absorptions at $\lambda \sim$ 4911\AA\  at  $\lambda \sim$ 5004\AA\  (partially filled up by  [OIII]  $\lambda$ 5007\AA ) and a previously unidentified absorption at $\lambda \sim$5183\AA .
We interpret these narrow absorptions as the FeII multiplet (42)  $\lambda\lambda$ 4923.93,5018.44,5169.03\AA, similar to those detected in a variety of transient luminous objects like the LBV candidate NGC~2366~V1 or SN type IIn  1995G, 1999el or 1999eb, blueshifted by the same amount as the Balmer absorption features  i.e.  $\sim$800 km/s to $\lambda\lambda$ 4911,5005,5155 \AA.  

The analysis of the photometric data provided by the CSS puts a strong upper limit to the possible variability of {\it PHL 293B\/}.
Basically any optical  yearly variability allowed by the data should be smaller than 0.02 magnitudes  at the 3 sigma level in the 8.5 years between April 2005 and September 2013. The possibility of  any secular trend in the luminosity of {\it PHL 293B\/}  in the last 15 years (considering only CCD data) is also limited to at most 
0.02 magnitudes at the 3 sigma level. 

This lack of variability and the observed strength of the Balmer and FeII absorptions rule out any transient of the type of an LBV or SN type IIn
as the origin of the blue shifted absorptions of H and FeII.

The evidence points to either a young and dense expanding supershell   or a stationary cooling wind, both  driven by the young cluster wind.
We suggest that the observed absorptions  and broad Balmer emissions are originated in one of these scenarios which seem capable of explaining the observed spectral features, the constant (within errors) photometric history and the rarity of the phenomenon.

Many starbursts, both nearby and at high z, show blueshifted far UV ISM narrow absorption lines. On that basis, coupled with the observational evidence, we expect that the far UV spectrum of {\it PHL 293B\/}  will show blue shifted ISM lines and moderate Ly$\alpha$ emission with perhaps a P~Cygni-like profile.

We have to bear in mind that there are not many HII galaxies known to show prominent broad wings in their emission lines and none with the quality of the data presented here for {\it PHL 293B\/}. This leaves open the possibility that we are witnessing an event not detected in other systems due to low S/N data.
Even if it is true that we haven't been actively looking for them, and that data of the quality and variety discussed here is not widely available,  it is still puzzling that Fe absorptions have not been detected in other star forming HII galaxies. A search should be performed in high dispersion, high signal-to-noise spectra of HII galaxies to investigate the presence of supershells in starbursts with or without strong broad components in the Balmer lines. 
Until such data is gathered and analysed, the fact that we do not see many HII galaxies showing spectra similar to  
{\it PHL 293B\/} means that this  may be a relatively short duration stage  in the evolution of compact and massive stellar clusters,
lasting perhaps only a few thousand years.

%This is probably the first detection of blue shifted optical absorption lines (JAJA)

\section*{Acknowledgments}

ET and RT are grateful to the Mexican Research Council (CONACYT) for support under  
 grants 2008-103365 and  2010-01-155046. Financial support for this work has also been provided by the Spanish \textit{Ministerios de Educaci\'{o}n y Ciencia and Ciencia e Innovaci\'on}  under grants AYA2007-67965-C03-03 and
AYA2010-21887-C04-03.   
Our team enjoyed the hospitality of the Institute of Astronomy, Cambridge, of the Departamento de F\'\i sica 
Te\'orica of the Universidad Aut\'onoma de Madrid and  of the Observatorio Astron\'omico of the Universidad Nacional de La Plata (Argentina) during  fruitful visits when this paper was started and developed. We thank Vahram Chavushyan  for suggesting the identification of the FeII absorption features in $\lambda\lambda$4911, 5155 \AA\ and discussions with Daniel Kunth helped to clarify the estimates of the analysis of these FeII features. Greatly enjoyed were discussions with Guillermo Tenorio-Tagle, Sergiy Silich and Sergio Mart\'\i nez-Gonz\'alez, from which a picture of the stationary wind scenario started to emerge.
We are grateful to Vital Fern\'andez for helping with the reduction of the WHT data. We thank the {\it X-shooter\/} data reduction team for valuable help and suggestions and an anonymous referee whose questionings greatly improved the clarity of the paper.

The CSS survey is funded by the National Aeronautics and Space
Administration under Grant No. NNG05GF22G issued through the Science
Mission Directorate Near-Earth Objects Observations Program.  The CRTS
survey is supported by the U.S.~National Science Foundation under
grants AST-0909182. This research has made use of the NASA/IPAC Extragalactic Database (NED) which is operated by the Jet Propulsion Laboratory, California Institute of Technology, under contract with the National Aeronautics and Space Administration.

\bsp

\label{lastpage}

\end{document}